\documentclass[prb, twocolumn, showpacs]{revtex4-1}
\usepackage{float}
\usepackage{bbm}
\usepackage{epsfig}
\usepackage{epstopdf}
\usepackage{graphicx}
\usepackage{amsmath,amssymb}
\usepackage{physics}
\usepackage{color}
\usepackage{hyperref}
\usepackage[caption=false]{subfig}
\usepackage{siunitx, booktabs}
\usepackage{diagbox, eqparbox, hhline}
\newcolumntype{P}[1]{>{\centering\arraybackslash}p{#1}}

\begin{document}

\title{Negative exchange interactions in coupled few-electron quantum dots}

\author{Kuangyin Deng$^1$}
\author{F.~A.~Calderon-Vargas$^1$}
\author{Nicholas J. Mayhall$^2$}
\author{Edwin Barnes$^1$}
\email{efbarnes@vt.edu}
\affiliation{$^1$Department of Physics, Virginia Tech, Blacksburg, VA 24061, USA\\ $^2$Department of Chemistry, Virginia Tech, Blacksburg, VA 24061, USA}

\begin{abstract}
It has been experimentally shown that negative exchange interactions can arise in a linear three-dot system when a two-electron double quantum dot is exchange coupled to a larger quantum dot containing on the order of one hundred electrons. The origin of this negative exchange can be traced to the larger quantum dot exhibiting a spin triplet-like rather than singlet-like ground state. Here, we show using a microscopic model based on the configuration interaction (CI) method that both triplet-like and singlet-like ground states are realized depending on the number of electrons. In the case of only four electrons, a full CI calculation reveals that triplet-like ground states occur for sufficiently large dots. These results hold for symmetric and asymmetric quantum dots in both Si and GaAs, showing that negative exchange interactions are robust in few-electron double quantum dots and do not require large numbers of electrons.
\end{abstract}


\maketitle

\section{Introduction}\label{sec:Intro}
Spins in GaAs and Si lateral quantum dots are promising candidates for implementing a quantum computer due to their scalability ~\cite{Loss1998,Hanson2007,Zwanenburg2013}, long coherence times ~\cite{Bluhm2010,Maune2012,Veldhorst2014,Malinowski2016}, and rapid gate operations ~\cite{Petta2005,Foletti2009,Medford2013a}. A qubit can be encoded in different ways using the spin states of one or more electrons trapped in one or more quantum dots, among these the single-spin ~\cite{Takeda2016a,Kawakami2016,Nowack2011,Veldhorst2014,Veldhorst2015a,Zajac2017}, singlet-triplet ~\cite{Petta2005,Maune2012,Foletti2009,Cerfontaine2014,Wu2014,Shulman2012,Nichol2016}, resonant exchange ~\cite{Laird2010,Medford2013a,Medford2013,Eng2015}, and hybrid spin ~\cite{Shi2012,Koh2012,Kim2014,Kim2015b,Cao2016} qubits have been successfully implemented in the laboratory. In many of these systems qubits are coupled to each other by means of the tunneling-based effective exchange interaction, which has the advantage of producing fast gates controlled electrically with gate voltages.~\cite{Petta2005,Maune2012,Brunner2011}. Furthermore, it has been recently demonstrated that symmetric exchange pulses substantially reduce the sensitivity of qubit gates to charge noise ~\cite{Hu2006,Reed2016,Martins2016,Barnes2016,Zhang2017,Yang2017,Yang2017a,Shim2017}.

Notwithstanding these advantages, the short-ranged nature of the exchange coupling ~\cite{Burkard1999,Li2009} is a potential hindrance towards scalability. However, this limitation can be circumvented by using an intermediate quantum system as a mediator ~\cite{Mehl2014c,Baart2016a,Mi2017}, for example a multielectron quantum dot ~\cite{Srinivasa2015,Croot2017,Malinowski_Thesis}. In this line, Refs.~\onlinecite{Martins2017,Malinowski2017a} study the spin properties of a multielectron GaAs quantum dot (with an estimated number of electrons between 50 and 100) exchange coupled to a single-electron quantum dot, which in turn is coupled to another single-electron quantum dot. This linear three-dot system is studied under magnetic fields both parallel and perpendicular to the two-dimensional electron gas (2DEG). In particular, the aforementioned works show that, at the transition between odd and even occupation number, the multielectron ground state is singlet-like for small hybridization and becomes triplet-like once the central electron has totally moved to the multielectron dot. As a result, the usually positive exchange energy becomes negative, even at zero magnetic field. This finding is not only important for understanding the properties of multielectron quantum dots, but also for performing dynamical decoupling on exchange-coupled spins. If the exchange coupling is restricted to be nonnegative, then special techniques are needed to dynamically correct for noise errors during gate operations ~\cite{Wang2012,Kestner2013,Wang2014}, which generally leads to longer gate times. Instead, if the exchange coupling can be tuned to both positive and negative values, then standard decoupling techniques can be used ~\cite{Goelman1989}, and this issue is avoided.

It has been demonstrated that negative exchange energy in a quantum dot with just two-electrons can be induced by a non-zero out-of plane magnetic field ~\cite{Wagner1992,Baruffa2010a,Zumbuhl2004,Mehl2014a}. Here, the out-of-plane magnetic field leads to a compression of the orbital wave functions and a larger electron-electron repulsion, which makes triplets energetically favorable. However, an in-plane or zero magnetic field does not create a wave-function compression, and thus it does not induce a negative exchange energy in a doubly occupied quantum dot, i.e. the ground state is always the singlet. In fact, there is a two-electron ground state theorem ~\cite{Lieb1962,2electron_theorem}, which states that in the absence of spin or velocity-dependent forces (the force exerted by the in-plane magnetic field is negligible since it is along the strong confinement perpendicular to the 2DEG) the state of lowest energy must be non-degenerate. An extension of the two-particle theorem to an arbitrary number of particles is given in Ref.~\onlinecite{Lieb1962}. This theorem correctly predicts the ground state of many electrons in a linear array, as shown in Ref.~\onlinecite{Riiser1993a}. Nonetheless, the multielectron ground state theorem does not apply to electrons interacting with central forces ~\cite{Lieb1962} and, for the multielectron quantum dot, the lower full orbitals do exert an effective central force onto higher orbitals. Therefore, there is no fundamental theorem or principle that prevents a triplet-like eigenstate from being the ground state of a multielectron quantum dot, regardless of the magnitude and direction of the magnetic field, as demonstrated in recent ~\cite{Martins2017,Malinowski2017a} and earlier ~\cite{Folk2001,Lindemann2002} experiments with multielectron quantum dots.
 
\indent In this work we demonstrate that a quantum dot does not require tens of electrons to exhibit negative exchange energy, instead, as few as 4 electrons are enough to have a triplet-like ground state in a quantum dot with zero magnetic field. We do this by performing a detailed numerical analysis employing the configuration interaction method with up to 14 electrons in both GaAs and Si quantum dots. Moreover, we use the full configuration interaction to determine the ground state of an elliptically shaped four-electron quantum dot with different eccentricities and, in doing so, we identify a threshold, in both GaAs and Si quantum dots, at which the exchange energy flips sign.

\indent The paper is divided in four sections. In Sec.~\ref{sec:Hubbard_model_Hund's_rule}, we use a simple Hubbard model to study and give a general picture of the system presented in Ref.~\onlinecite{Martins2017}. Then, in Sec.~\ref{sec:configuration_interaction}, we use a configuration interaction method to determine the ground state of a multielectron quantum dot with parabolic potential, where we consider different number of electrons and dot sizes. Moreover, a full configuration interaction calculation shows that for four electrons the ground state is triplet-like for all the dot sizes we considered. Finally, in Sec.~\ref{sec:elliptical_potential}, we use the full configuration interaction method to calculate the exact eigenenergies of four electrons confined in an elliptically shaped quantum dot, showing the effect of the dot asymmetry in the occurrence of triplet-like ground states.

\section{Hubbard model and Hund's rule}\label{sec:Hubbard_model_Hund's_rule}
\begin{figure}[!tbp]
\centering
\includegraphics[trim=0cm 10cm 0cm 9cm, clip=true,width=8.5cm, angle=0]{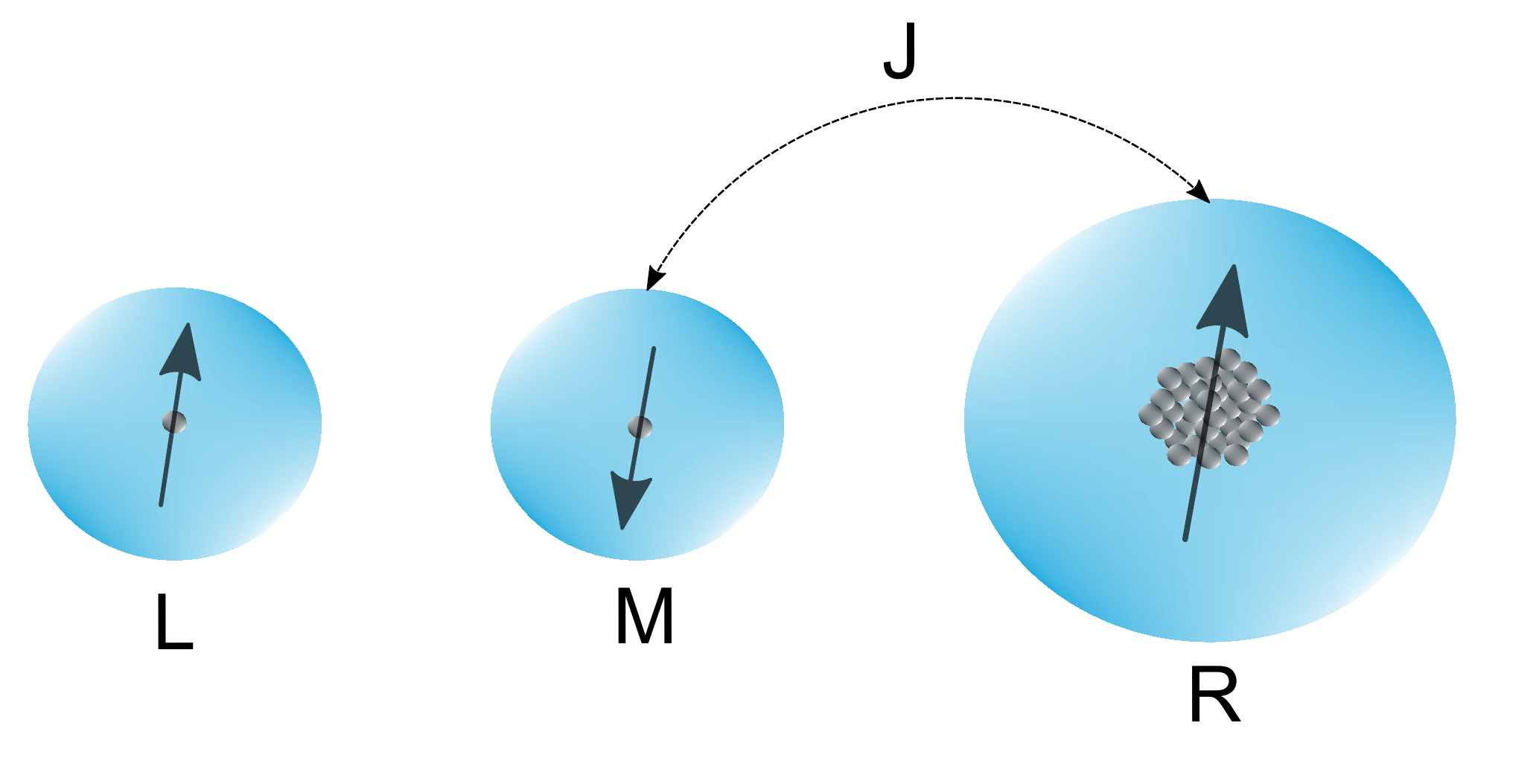}
\caption{Illustration of the three dot system, where the $L$ and $M$ dots form a two-electron double quantum dot and $R$ is the multielectron quantum dot. In the main text, we analyze the effective exchange interaction, $J$, between the middle ($M$) and right ($R$) quantum dots.}\label{Fig.0}
\end{figure}

\begin{figure}[!tbp]
\centering 
\includegraphics[trim=0cm 0cm 0cm 0cm, clip=true,width=8.7cm, angle=0]{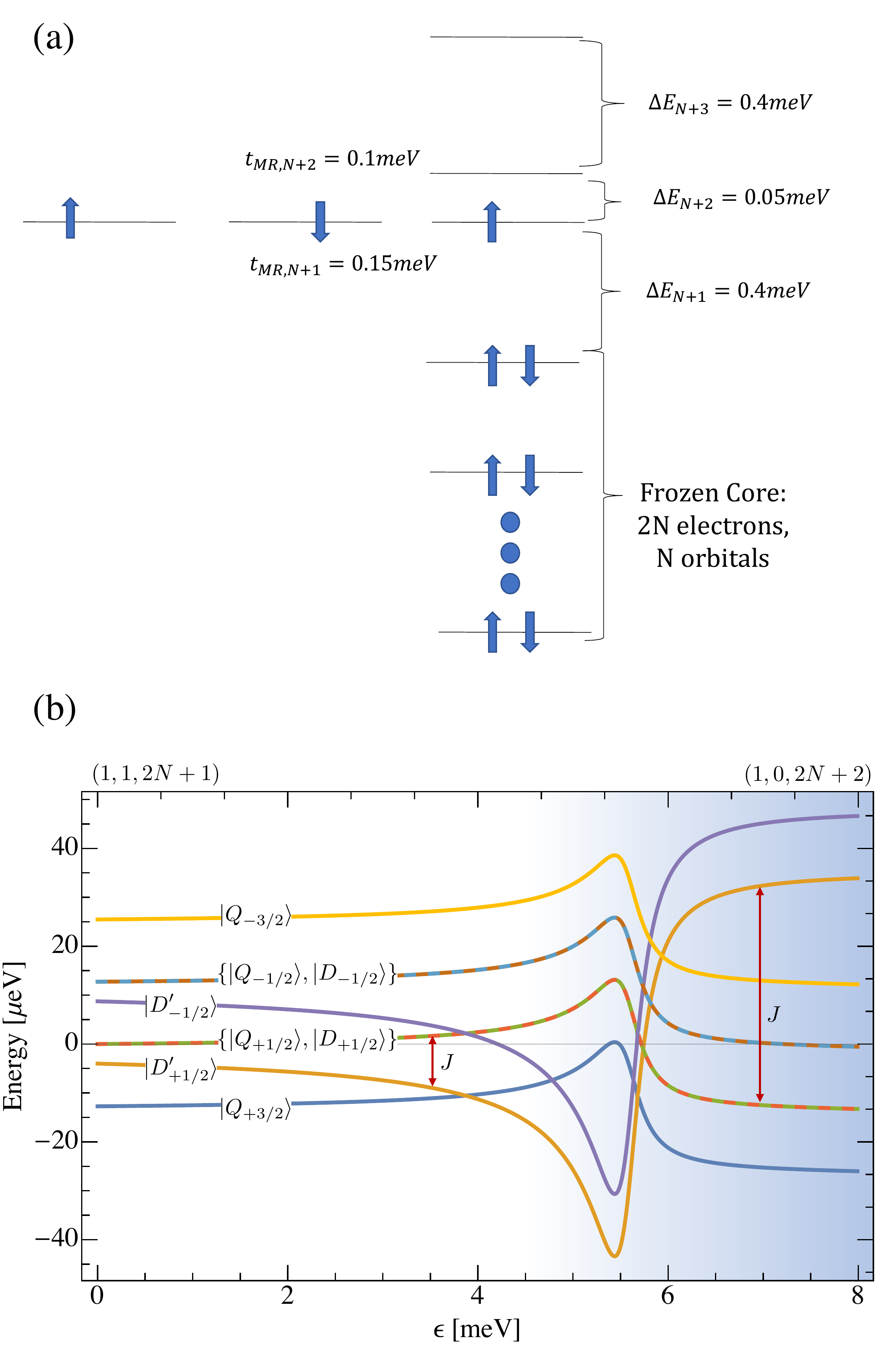}
\caption{(a) Schematic of the energy levels and charge configuration for the three-dot system. Here, $\Delta E_{i}$ is the difference between two orbitals on the multielectron quantum dot. The schematic only shows the tunnel coupling between the middle and right dot, where $t_{MR,N+1}$  and $t_{MR,N+2}$ are the tunneling amplitudes between the middle dot's single-orbital and the lowest two orbitals above the frozen core in the multielectron quantum dot. (b) Eigenenergy spectrum, calculated as a function of the detuning $\epsilon$, for the three-dot system at the transition between $(1,1,2N+1)$ and $(1,0,2N+2)$ charge configurations. The eigenstates $\ket{D_{+1/2}}$ ($\ket{D_{-1/2}}$) and $\ket{Q_{+1/2}}$ ($\ket{Q_{-1/2}}$) are almost degenerate and the exchange energy $J$ is the difference between the triplet-like state $\ket{D_{\pm1/2}}$ and the singlet-like state $\ket{D'_{\pm1/2}}$.}\label{fig:1}
\end{figure}

We start our analysis with a simple Hubbard model that describes the system studied in Ref. \onlinecite{Martins2017}, i.e. a multielectron quantum dot (rightmost) with $2N+1$ electrons ($N=50$) tunnel-coupled to a double quantum dot containing two electrons, see Fig.~\ref{Fig.0}. We keep as many orbitals (single-particle energy levels) as necessary in the right dot and only one orbital in each of the other two quantum dots (see Fig.~\hyperref[fig:1]{\ref*{fig:1}(a)}). The system's Hamiltonian is
\begin{equation}\label{eq:Hubbard_model}
H=H_0+H_z+H_A+H_U,
\end{equation}
where
\begin{align}
H_0=&\sum_\sigma[\sum_l\epsilon_{R,l}n_{R,l\sigma}+\epsilon_Mn_{M,\sigma}+ \epsilon_Ln_{L,\sigma}\nonumber\\
&-\sum_l t_{MR,l}(c_{R,l\sigma}^\dagger c_{M,\sigma}+c_{M,\sigma}^\dagger c_{R,l\sigma})\nonumber\\
&-t_{LM}(c_{M,\sigma}^\dagger c_{L,\sigma}+c_{L,\sigma}^\dagger c_{M,\sigma})],\\
H_Z=&\frac{E_B}{2}[\sum_l(n_{R,l\uparrow}-n_{R,l\downarrow})+n_{M,\uparrow}-n_{M,\downarrow}+n_{L,\uparrow}-n_{L,\downarrow}],\\
H_A=&A_R\sum_l(c_{R,l\uparrow}^\dagger c_{R,l\downarrow}+c_{R,l\downarrow}^\dagger c_{R,l\uparrow})+A_{L,M}(c_{M,\uparrow}^\dagger c_{M,\downarrow}\nonumber\\
&+c_{M,\downarrow}^\dagger c_{M,\uparrow}+c_{L,\uparrow}^\dagger c_{L,\downarrow}+c_{L,\downarrow}^\dagger c_{L,\uparrow}),\\
H_U=&\sum_lU_{R,l}n_{R,l\uparrow}n_{R,l\downarrow}+\sum_{l_1\ne l_2, \sigma,\sigma'}U_{R,l_1l_2}n_{R,l_1\sigma}n_{R,l_2\sigma'}\nonumber\\
&+U_Mn_{M,\uparrow}n_{M,\downarrow}+U_Ln_{L,\uparrow}n_{L,\downarrow}\nonumber\\
&+\sum_{l_1\ne l_2, \sigma,\sigma'}U_{R,l_1l_2 l_2 l_1}c_{R,l_1\sigma}^\dagger c_{R,l_2\sigma'}^\dagger c_{R,l_1\sigma'}c_{R,l_2\sigma}.\label{eq:Coulomb_onsite_interaction_exchange}
\end{align}
Here, $n_{\alpha,l \sigma}=c^{\dagger}_{\alpha,l \sigma}c_{\alpha,l \sigma}$ is the number operator for the single-particle states in the left ($L$), middle ($M$), and right ($R$) quantum dot ($\alpha=L,M,R$; $\sigma=\uparrow,\downarrow$; and $l$ is the right dot's $l$-th single-particle state), $\epsilon_{\alpha,l}$ denotes the single-particle energies, $t_{\alpha\beta} $ is the tunneling amplitude between dots ($\alpha,\beta=L,M,R$), $E_B$ is the Zeeman energy, $A_\alpha$ is the hyperfine interaction (proportional to the dot size) between electrons and the nuclear spin bath, $U_\alpha$ is the ``on-site'' Coulomb interaction, and finally, $U_{R,l_1 l_2} $  and $U_{R,l_1l_2 l_2 l_1}$ are the Coulomb interaction and exchange term between orbitals $l_1$ and $l_2$ in the right dot, respectively. The large number of electrons in the right dot makes the numerical calculation of the eigenenergies too difficult to carry out without any sort of approximation. Accordingly, we make use of the so-called ``frozen-core'' approximation (FCA), where we keep the right dot's $2N$ core electrons in the lowest non-interacting states and only allow a valence electron to occupy higher energy levels. It is worth mentioning that, due to the large number of core electrons, we are only considering the direct Coulomb interaction between the core and the valence electrons, which causes a general energy shift. In the following sections we will consider cores with fewer electrons, and thus the FCA will also take into account the Coulomb exchange interaction between core and valence electrons.\\
\indent In atomic physics, it is well known that in certain configurations a combination of electron-electron repulsion and electron-nucleus attraction makes high-spin states energetically more favorable than any other lower-spin state arising from the same configuration ~\cite{Boyd1984}; this is commonly known as Hund's multiplicity rule. Similarly, for the multielectron quantum dot the exchange term,  $U_{R,l_1l_2 l_2 l_1}$, in Eq.~\eqref{eq:Coulomb_onsite_interaction_exchange} induces magnetic correlations among the electron spins and, as a result, it lowers the energy of the eigenstates with spin 1. This is more evident if we set $J_{R,l_1l_2 }^F \equiv U_{R,l_1l_2 l_2 l_1}$ and, using Pauli matrix identities, we rewrite the exchange energy in Eq.~\eqref{eq:Coulomb_onsite_interaction_exchange} as~~\cite{CMFT_Alexander}
\begin{align}
&\sum_{l_1\ne l_2, \sigma ,\sigma'}U_{R,l_1l_2 l_2 l_1}c_{R,l_1\sigma}^\dagger c_{R,l_2\sigma'}^\dagger c_{R,l_1\sigma'}c_{R,l_2\sigma}\nonumber\\
=&-2\sum_{l_1\ne l_2}J_{R,l_1l_2}^F (\mathbf{S}_{R,l_1}\cdot\mathbf{S}_{R,l_2}+\frac{1}{4}n_{Rl_1} n_{Rl_2}),
\end{align}
where $\mathbf{S}_{R,l_i}$ is the spin operator acting on the $l_i$-th single-particle state.\\
\indent Since we assume that the $2N$ core electrons in the right dot are ``frozen'' (FCA), we are effectively dealing with a three-electron system. The spin Hamiltonian for three electrons coupled by nearest-neighbor exchange interactions and subject to a magnetic field is $H'=J_{LM}\left(\mathbf{S}_L\cdot\mathbf{S}_M-1/4\right)+J_{MR}\left(\mathbf{S}_M\cdot\mathbf{S}_R-1/4\right)-E_B(S_{z,L}+S_{z,M}+S_{z,R})$, where $J_{\alpha\beta}$ acts as an effective exchange interaction and $E_B$ is the Zeeman energy. The eight spin eigenstates of this Hamiltonian form a quadruplet $Q$,
\begin{align}
\ket{Q_{+3/2}}=&\ket{\uparrow\uparrow\uparrow},\\
\ket{Q_{+1/2}}=&\frac{1}{\sqrt{3}}(\ket{\downarrow\uparrow\uparrow}+\ket{\uparrow\downarrow\uparrow}+\ket{\uparrow\uparrow\downarrow}),\\
\ket{Q_{-1/2}}=&\frac{1}{\sqrt{3}}(\ket{\uparrow\downarrow\downarrow}+\ket{\downarrow\uparrow\downarrow}+\ket{\downarrow\downarrow\uparrow}),\\
\ket{Q_{-3/2}}=&\ket{\downarrow\downarrow\downarrow},
\end{align}
and high- and low-energy doublets, which, in the absence of tunneling between left and middle dots, have the following simple form
\begin{align}
\ket{D_{+1/2}}=&\frac{1}{\sqrt{6}}(-2\ket{\downarrow\uparrow\uparrow}+\ket{\uparrow\downarrow\uparrow}+\ket{\uparrow\uparrow\downarrow}),\\
\ket{D_{-1/2}}=&\frac{1}{\sqrt{6}}(-2\ket{\uparrow\downarrow\downarrow}+\ket{\downarrow\uparrow\downarrow}+\ket{\downarrow\downarrow\uparrow}),\\
\ket{D'_{+1/2}}=&\frac{1}{\sqrt{2}}(\ket{\uparrow\uparrow\downarrow}-\ket{\uparrow\downarrow\uparrow}),\\
\ket{D'_{-1/2}}=&\frac{1}{\sqrt{2}}(\ket{\downarrow\downarrow\uparrow}-\ket{\downarrow\uparrow\downarrow}).
\end{align}
Here, the spin eigenstates $\ket{D_{+1/2}}$ ($\ket{D_{-1/2}}$) and $\ket{Q_{+1/2}}$ ($\ket{Q_{-1/2}}$) are almost degenerate, with an energy $E_{D_{\pm 1/2}}=\mp E_B/2$, whereas the low-energy doublets have an energy $E_{D'_{\pm 1/2}}=-J_{MR}\mp E_B/2$. Therefore, the effective exchange energy between the middle and right dots is given by
\begin{equation}
J_{MR}=E_{D_{+ 1/2}}-E_{D'_{+ 1/2}},
\end{equation}
where $\ket{D_{+1/2}}$ and $\ket{D'_{+1/2}}$ are the lowest spin triplet-like and singlet-like eigenstates, respectively.\\
\indent We use the Hubbard model, Eq.~\eqref{eq:Hubbard_model}, to calculate the energies of the aforementioned spin eigenstates. To that end, we choose a set of parameters such that the resulting energy spectrum resembles the one reported in Ref.~\onlinecite{Martins2017}. Accordingly, we set the magnitude of the energy level splittings as $\Delta E_{N+1}=0.4\mathrm{meV}$, $\Delta E_{N+2}=0.05\mathrm{meV}$, $\Delta E_{N+3}=0.4\mathrm{meV}$, where $\Delta E_{l}=\epsilon_{R,l}-\epsilon_{R,l-1}$ and $\epsilon_{R,l}$ is the right dot's $l$-th single-particle energy level (see Fig.~\hyperref[fig:1]{\ref*{fig:1}(a)}). The tunneling amplitudes are $t_{MR,N+1}=0.15\mathrm{meV}$, $t_{MR,N+2}= 0.1\mathrm{meV}$, and $t_{LM}=0.02\mathrm{meV}$. The hyperfine interactions for the multielectron  quantum dot and double quantum dot are $A_R=0.4\mathrm{neV}$ and $A_{LM}=4\mathrm{neV}$, respectively. Finally, the ferromagnetic exchange term in the right dot is set equal to $J_{R,l_1l_2 }^F=0.1\mathrm{meV}$ and the in-plane magnetic field is $B=500\mathrm{mT}$. With these parameters we plot in Fig.~\hyperref[fig:1]{\ref*{fig:1}(b)} the eigenenergies as a function of the detuning between left and right dot, $\epsilon\equiv(\epsilon_M-\epsilon_{R,N+1})/2$. Notice that in the $(1,1,2N+1)$ configuration, where $(n_L,n_M,n_R)$ represents the number of electrons in the left, middle, and right dots, the singlet-like eigenstate $\ket{D'_{+1/2}}$ has lower energy than $\ket{D_{+1/2}}$ ($J_{MR}>0$), but as soon as the middle dot's electron tunnels into the right dot the exchange energy becomes negative, i.e. $E_{D_{+1/2}}<E_{D'_{+1/2}}$. The latter is caused by the right dot's ferromagnetic exchange term $J_{R,l_1l_2 }^F$, which lowers the energies of all two-electron states in the right dot with total spin equal to 1. This is analogous to Hund's rule in atomic physics.\\
\indent Thus far, we have shown that, by choosing the appropriate parameters, the simple Hubbard model qualitatively reproduces the experimental results presented in Ref.~\onlinecite{Martins2017}. Moreover, the model shows that the negative exchange energy is caused by the multielectron quantum dot exhibiting a spin triplet-like rather than singlet-like ground state. In this line, Ref.~\onlinecite{Malinowski2017a} presents comparable results using a Hubbard model similar to ours, with the difference that ours takes into account the hyperfine interaction (between electrons and the nuclear spin bath) and the Coulomb interaction between orbitals in the right dot. Nonetheless, the Hubbard model does not provide enough insight into the characteristics (minimum number of electrons, dot shape, etc.) a system must have in order to display a negative exchange energy under zero magnetic field. This is addressed below using a microscopic description of the multielectron quantum dot.
\section{Configuration Interaction for a multielectron quantum dot}\label{sec:configuration_interaction}
\subsection{ Frozen-core approximation}
Here we determine the ground state of a multielectron quantum dot using a configuration interaction (CI) approach. The large number of electrons forces us to use, once again, the ``frozen-core'' approximation, which now also takes into account the Coulomb exchange interaction between core and valence electrons.\\
\indent The system can be described by the valence effective Hamiltonian ~\cite{Durand1975,Malrieu2014}:
\begin{align}\label{eq:valence_effective_Hamiltonian}
H_v=&[E_v+\sum_i^{core}h_{ii}+\sum_{i<j}^{core}(ii|jj)-(ij|ji)]\nonumber\\
&+\sum_{r,s}^{val}\tilde{h}_{rs}c_r^\dagger c_s+\frac{1}{2}\sum_{p,q,r,s}^{val}{(pq|rs)c_p^\dagger c_r^\dagger c_s c_q},
\end{align}
with
\begin{equation}\label{eq:molecular_orbital_interaction}
(kl|mn)=\int{\phi_k^*(r_1)\phi_l(r_1)\frac{1}{r_{12}}\phi_m^*(r_2)\phi_n(r_2)}dr_1 dr_2,
\end{equation}
where the single-particle orbitals are denoted by $\phi_\alpha$, and the summations marked ``core'' and ``val'' are over orbitals occupied by core or valence electrons, respectively. The terms inside the bracket in Eq.~\eqref{eq:valence_effective_Hamiltonian}, which comprises the total single-particle energy of the valence electrons ($E_v$) and the core's energy, add up to a constant, and thus they only shift the energy scale. The energy term $\tilde{h}_{rs}$ is defined as
\begin{equation}\label{eq:hopping_exchange}
\tilde{h}_{rs}=t_{rs}+\sum_i^{core}{[(rs|ii)-(ri|is)]},
\end{equation}
where $t_{rs}$ is the electron hopping between valence orbitals, and $(rs|ii)$ and $(ri|is)$ are the Coulomb interaction and exchange coupling between the valence and core electrons, respectively.\\
\indent In our numerical analysis we consider $2N+2$ electrons ($2N$ core electrons and 2 valence electrons, $0\leq N\leq6$) living in 12 orbitals. We model the lateral gate confinement of the multielectron quantum dot with a symmetric parabolic potential. Thus, the appropriate single-particle orbitals are the eigenstates of the Fock-Darwin Hamiltonian
\begin{equation}\label{eq:Fock-Darwin_Hamiltonian}
H=\frac{1}{2m^*}(-i\hbar\nabla+\frac{e}{c}\mathbf{A})^2+\frac{1}{2}m^*\omega^2r^2,
\end{equation}
where $m^*$ is the effective electron mass, $\hbar\omega$ is the quantum dot confinement energy, and $r=\sqrt{x^2+y^2}$ is the radius. Following a numerical method developed in a previous work ~\cite{Barnes2011}, and setting the external magnetic field to zero, we determine the eigenenergies and ground eigenstate of the valence effective Hamiltonian. We perform this calculation for both GaAs and Si quantum dots (see Table~\ref{Tab:1}); using for GaAs (Si) the effective electron mass $m^*=0.067m_e$ ($m^*=0.19m_e$), where $m_e$ is the electron mass, and the dielectric constant of the host material $\kappa=13.1\epsilon_0$ ($\kappa=11.68\epsilon_0$). In contrast to GaAs, Si quantum dots present a two-fold degenerate ground state. This valley degeneracy can be lifted and finely tuned by an out-of-plane electric field ~\cite{Yang2013,Veldhorst2014}. Here, we assume that such techniques have been employed to achieve a sufficiently large valley splitting, and thus we do not include a valley coupling parameter in our calculations. A comprehensive analysis including valley effects would require a detailed microscopic understanding of the intervalley coupling, which is likely device specific and is beyond the scope of this paper.

Before discussing our results, it is important to note that the multielectron quantum dot has shells at $\eta=2, 6, 12, 20, \ldots, (n^2+3n+2)$, where $\eta$ is the total number of electrons in the dot and $n=0, 1, 2, 3, \ldots$ is the principal quantum number. This is a consequence of the ($n+1$)-fold degeneracy of the quantum dot's eigenenergies, which stems from the dot's effective confinement having a symmetry very close to circular ~\cite{Reimann2002}. We are primarily interested in situations where the multielectron quantum dot can be used as a spin qubit, i.e., it initially contains an odd number of electrons where all but one electron completely fill a number of shells and form a spin singlet-like state, leaving a net spin $1/2$ from the remaining unpaired electron ~\cite{Barnes2011,Nielsen2013a,Higginbotham2014}. To examine the sign of the exchange interaction with a neighboring single-electron quantum dot we consider that the neighboring electron tunnels into the large quantum dot, giving now two valence electrons in that dot. We therefore consider first the simplest case involving just a pair of electrons, and then for larger number of electrons we focus on electron numbers $\eta= 4, 8, 14, \ldots, (n_{max}^2 +3n_{max} +4)$,  which correspond to two valence electrons and where $n_{max}$ is the principal quantum number of the highest full shell in the core. 

Our results, presented in Table~\ref{Tab:1}, show that when the multielectron quantum dot contains only two electrons, only a singlet ground state can be realized; this is in accordance with the two-electron ground state theorem ~\cite{Lieb1962,2electron_theorem}. Incidentally, in the case of valley degeneracy or near-degeneracy, a pair of electrons in a Si quantum dot would not necessarily follow the aforementioned theorem since electrons in different valleys could be treated as different species ~\cite{Hada2003}, which would violate the theorem's assumption that the potential is symmetric under permutations ~\cite{Lieb1962}. In Table~\ref{Tab:1} we also see that, for more than two electrons, triplet states are possible depending on the size of the dot. This indicates that having a core of electrons completely occupying lower energy orbitals is important for creating a triplet ground state. The fact that, at least for $\eta>4$, whether the ground state is a singlet or a triplet depends on the size of the dot suggests that the orbital spacing plays an important role. We know that the energy difference between shells is inversely proportional to the square radius of the quantum dot, so that small dots present well defined energy gaps between shells. Consequently, for a pair of valence electrons above a full shell and for a sufficiently large energy gap between shells (small dot), Coulomb interactions between valence and core electrons are likely reduced since excitations from core to valence orbitals are suppressed. This picture is consistent with what we observe in Table~\ref{Tab:1}, where for sufficiently small dots ($\omega/\omega_0\geq4$ for GaAs and $\omega/\omega_0\geq16$ for Si, where $\hbar\omega_0=1.0 \mathrm{meV}$) the ground state is always triplet-like, while for bigger dots ($\omega/\omega_0\leq2$ for GaAs and $\omega/\omega_0\leq8$ for Si) the comparatively larger Coulomb interactions between valence and core electrons increases the likelihood of singlet-like ground states.
\begin{table}[tbp]
\centering
\caption{Ground states (S=Singlet and T=Triplet) for different dot sizes and number of electrons. Here, $\hbar\omega_0=1.0 \mathrm{meV}$ for both GaAs and Si quantum dots.}\label{Tab:1}
\subfloat[Ground states table for GaAs.]{\begin{tabular}{@{}|P{2.5cm}||@{} P{0.5cm}| @{}P{0.7cm}@{}|@{}P{0.7cm}@{}|@{}P{0.7cm}@{}|@{}P{0.7cm}@{}|@{}P{0.7cm}@{}|@{}P{0.7cm}@{}|@{}}
  \hhline{-||-------}
  \diagbox[width=\dimexpr\eqboxwidth{wd} + 8.8\tabcolsep\relax, height=1cm]{{\scriptsize \# of electrons}}{\raisebox{-1ex}{$\omega/\omega_0$}}
  &~ 0.25 & 0.5 & 1 & 2 & 4 &8 &16\\
  \hhline{=::=======}
  \eqmakebox[wd]{2} & ~~ S &S & S & S & S & S & S \\
  \hhline{-||-------}
  \eqmakebox[wd]{4} & ~~ T &T & T & T & T & T & T \\
  \hhline{-||-------}
  \eqmakebox[wd]{8} & ~~ S &T & T & T & T & T & T\\
  \hhline{-||-------}
  \eqmakebox[wd]{14} & ~~ S &S & S & S & T & T & T \\
  \hhline{-||-------}
\end{tabular}}
\bigskip
\\
\subfloat[Ground states table for Si.]{\begin{tabular}{@{}|P{2.5cm}||@{} P{0.5cm}| @{}P{0.7cm}@{}|@{}P{0.7cm}@{}|@{}P{0.7cm}@{}|@{}P{0.7cm}@{}|@{}P{0.7cm}@{}|@{}P{0.7cm}@{}|@{}}
  \hhline{-||-------}
  \diagbox[width=\dimexpr\eqboxwidth{wd} + 8.8\tabcolsep\relax, height=1cm]{{\scriptsize \# of electrons}}{\raisebox{-1ex}{$\omega/\omega_0$}}
  &~ 0.25 & 0.5 & 1 & 2 & 4 &8 &16\\
  \hhline{=::=======}
  \eqmakebox[wd]{2} & ~~ S &S & S & S & S & S & S \\
  \hhline{-||-------}
  \eqmakebox[wd]{4} & ~~ T &T & T & T & T & T & T \\
  \hhline{-||-------}
  \eqmakebox[wd]{8} & ~~ S &S & S & T & T & T & T\\
  \hhline{-||-------}
  \eqmakebox[wd]{14} & ~~ S &S & S & S & S & S & T \\
  \hhline{-||-------}
\end{tabular}}
\end{table}
\subsection{Full Configuration interaction for a four-electron dot}
The ``frozen-core'' approximation (FCA) was instrumental in the calculation of the results presented in Table \ref{Tab:1} and, therefore, it is important to probe the accuracy of this approximation. To that end, we use the full configuration interaction (full CI) method to calculate the exact eigenenergies of a four-electron dot with variable size and zero magnetic field. In the numerical calculation we consider 4 electrons living in the 10 lowest orbitals of a parabolic potential. Our results show that the ground state of this system is triplet-like regardless of the dot size, in accordance with the results obtained through the FCA. We also notice that the higher-orbital-content of the ground state increases proportionally to the dot size. This is due to the reduction in the energy gap between orbitals when the size of the dot increases, which allows the mixing with higher orbitals. In this regard, the FCA only provides an estimation of the ground state's orbital content, and thus the FCA's accuracy is expected to diminish for large-size dots. Nonetheless, the FCA remains a good approximation within the dot size range considered in this work.
\section{Full Configuration interaction with Elliptical potential}\label{sec:elliptical_potential}
\begin{figure}[!tbp]
\centering
\includegraphics[trim=0cm 5cm 0cm 5cm, clip=true,width=8.5cm, angle=0]{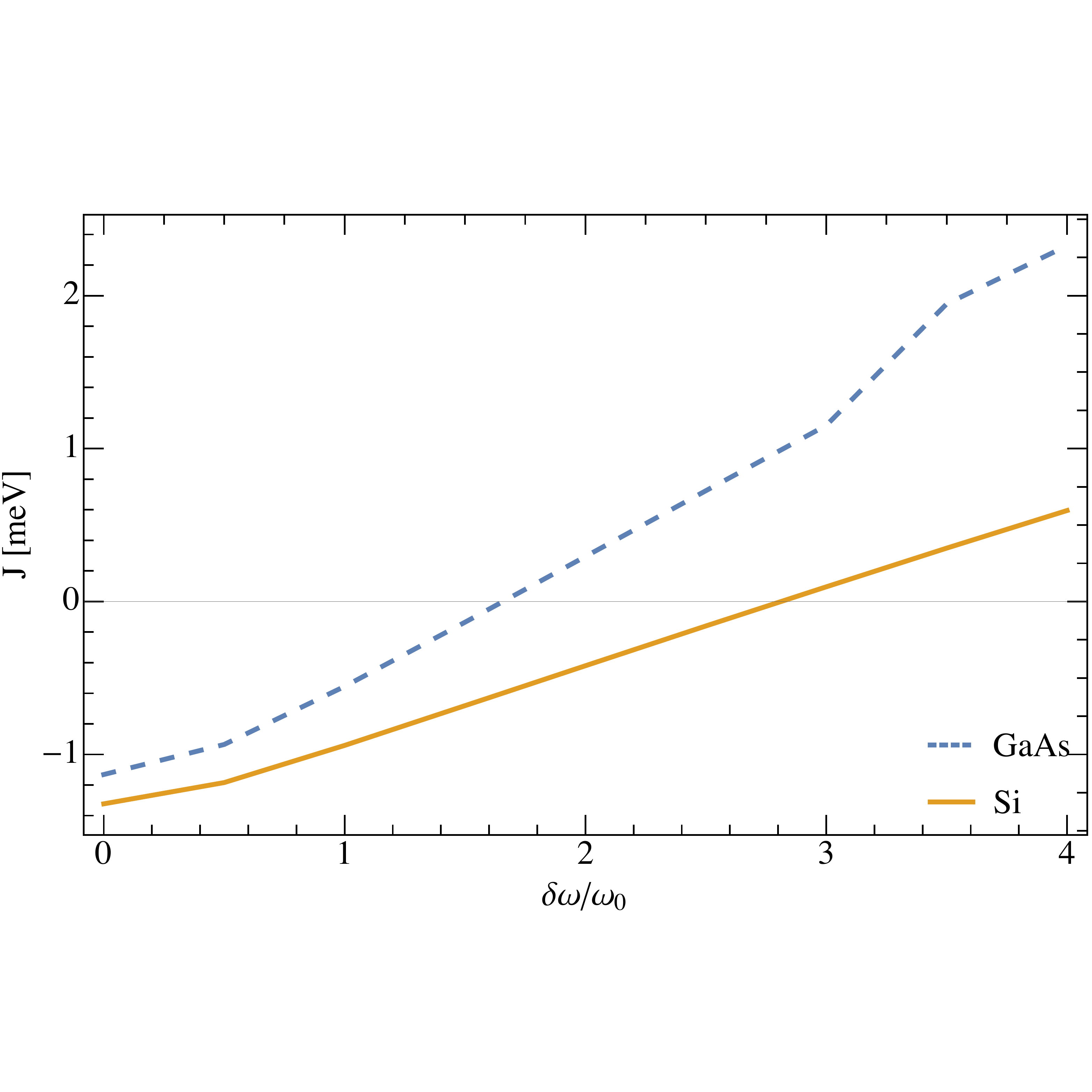}
\caption{Exchange energy vs. frequency difference of the elliptical potential.}\label{Fig.3}
\end{figure}
Apart from the dot size and number of electrons confined in a quantum dot, here we show that the shape of the dot also determines the occurrence of a triplet-like ground state. To that end we use, once again, the full CI method to calculate the exact eigenenergies of a four-electron quantum dot with elliptical potential and, in doing so, we show the effect of asymmetry on the exchange energy's magnitude and sign. In our calculation we consider 4 electrons residing in 10 orbitals, with a single-particle Hamiltonian given by
\begin{equation}
H=\frac{1}{2m^*}(-i\hbar\nabla+\frac{e}{c}\mathbf{A})^2+\frac{1}{2}m^*(\omega_1^2x^2+\omega_2^2y^2),
\end{equation}
where $\omega_1$ and $\omega_2$ are the frequencies of the harmonic oscillators for the $x$ and $y$ directions, respectively. In the absence of an external magnetic field, $\mathbf{B}=0$, the single-particle energies are the eigenvalues of the anisotropic two-dimensional harmonic oscillator:
\begin{equation}
\epsilon_{n_x,n_y}=\frac{1}{2}(\hbar\omega_1+\hbar\omega_2)+n_x \hbar\omega_1+n_y\hbar\omega_2,
\end{equation}
where we define $\tilde{n}=n_x+n_y$. Here, it is evident that the splitting between single-particle energies can be tuned by the frequency difference
\begin{equation}
\delta\omega=\omega_2-\omega_1,
\end{equation}
which effectively changes the eccentricity of the dot's elliptical potential. Accordingly, we calculate the eigenenergies of the four electrons (in both GaAs and Si quantum dots with zero magnetic field) using different magnitudes for $\delta\omega$, where, for convenience, we set $\frac{\hbar(\omega_2+\omega_1)}{2}=8\hbar\omega_0=8 meV$ and $\omega_2\geq\omega_1$. Our results are summarized in Fig.~\ref{Fig.3}, which shows  the exchange energy $J$ (given by the difference in energy between the lowest triplet-like and singlet-like eigenstates) as a function of the frequency difference $\delta\omega$. Notice that when $\delta\omega=0$, i.e. the potential is parabolic, $\tilde{n}$ becomes the principal quantum number and the principal energy levels (electron shells) corresponding to $\tilde{n}>0$ are degenerate. Here, the electron distribution is such that the lowest shell is full and the next degenerate shell contains two electrons. In this configuration, as shown in the previous section, the ground state is triplet-like. However, for non-zero $\delta\omega$ the degeneracy is lifted and, in our particular case, when $\delta\omega$ is greater than a certain threshold $\delta\tilde{\omega}$ ($\delta\tilde{\omega}\approx1.65\omega_0$ for GaAs and $\delta\tilde{\omega}\approx2.8\omega_0$ for Si) the split between the formerly degenerate orbitals with $\tilde{n}=1$ is large enough to favor a singlet-like ground state and, therefore, a positive exchange energy. A similar effect is observed with larger numbers of electrons and/or larger dot sizes (see Table \ref{Tab:1}).
\section{Conclusions}
In this work we have studied the conditions under which negative exchange interactions can occur in coupled few-electron quantum dots. The negative exchange interaction between a multielectron quantum dot (with odd occupation number) and a single-electron quantum dot (which in turn is coupled to a second single-electron quantum dot) has its roots in the larger quantum dot exhibiting a spin triplet-like ground state, which occurs once the smaller dot's electron has tunneled into the larger dot. This was demonstrated using a Hubbard model for a linear three-dot system ~\cite{Note} that reproduces the experimental results presented in Ref.~\onlinecite{Martins2017}, where negative exchange was observed. The larger quantum dot with an even number of electrons and zero magnetic field was further studied using a microscopic model based on the configuration interaction (CI) method with which we determined the ground state of the multielectron quantum dot. In this CI calculation we considered different combinations of total number of electrons and dot sizes (parabolic potential), showing that the occurrence of both triplet-like and singlet-like ground states depend on those parameters and that 4 electrons is the minimum needed to have a triplet-like ground state in both Si and GaAs quantum dots. Moreover, the effect of dot asymmetry on the exchange energy is also addressed via a full CI calculation of the energy spectrum for a four-electron quantum dot with elliptical potential. The full CI calculation is repeated for different eccentricities, revealing a threshold, in both GaAs and Si dots, at which the exchange energy flips signs. Future work will explore the equally interesting three-dot system where a multielectron quantum dot acts as a quantum mediator between two single-electron quantum dots. For now, the results presented in this work show that negative exchange interactions are robust in few-electron double quantum dots, and that all the potential advantages a tunable exchange interaction can provide are accessible with as few as 4 electrons in a double quantum dot. This is fundamental for scalability purposes since it avoids the need of large quantum dots, it prevents unwanted capacitive coupling between remote dots, and it enables simpler and faster dynamically corrected gate operations.

\section{Acknowledgments}
We thank Ferdinand Kuemmeth for helpful discussions. This work is supported by the Army Research Office (W911NF-17-0287).
\bibliographystyle{apsrev4-1}

\bibliography{library}

\begin{thebibliography}{70}%
\makeatletter
\providecommand \@ifxundefined [1]{%
 \@ifx{#1\undefined}
}%
\providecommand \@ifnum [1]{%
 \ifnum #1\expandafter \@firstoftwo
 \else \expandafter \@secondoftwo
 \fi
}%
\providecommand \@ifx [1]{%
 \ifx #1\expandafter \@firstoftwo
 \else \expandafter \@secondoftwo
 \fi
}%
\providecommand \natexlab [1]{#1}%
\providecommand \enquote  [1]{``#1''}%
\providecommand \bibnamefont  [1]{#1}%
\providecommand \bibfnamefont [1]{#1}%
\providecommand \citenamefont [1]{#1}%
\providecommand \href@noop [0]{\@secondoftwo}%
\providecommand \href [0]{\begingroup \@sanitize@url \@href}%
\providecommand \@href[1]{\@@startlink{#1}\@@href}%
\providecommand \@@href[1]{\endgroup#1\@@endlink}%
\providecommand \@sanitize@url [0]{\catcode `\\12\catcode `\$12\catcode
  `\&12\catcode `\#12\catcode `\^12\catcode `\_12\catcode `\%12\relax}%
\providecommand \@@startlink[1]{}%
\providecommand \@@endlink[0]{}%
\providecommand \url  [0]{\begingroup\@sanitize@url \@url }%
\providecommand \@url [1]{\endgroup\@href {#1}{\urlprefix }}%
\providecommand \urlprefix  [0]{URL }%
\providecommand \Eprint [0]{\href }%
\providecommand \doibase [0]{http://dx.doi.org/}%
\providecommand \selectlanguage [0]{\@gobble}%
\providecommand \bibinfo  [0]{\@secondoftwo}%
\providecommand \bibfield  [0]{\@secondoftwo}%
\providecommand \translation [1]{[#1]}%
\providecommand \BibitemOpen [0]{}%
\providecommand \bibitemStop [0]{}%
\providecommand \bibitemNoStop [0]{.\EOS\space}%
\providecommand \EOS [0]{\spacefactor3000\relax}%
\providecommand \BibitemShut  [1]{\csname bibitem#1\endcsname}%
\let\auto@bib@innerbib\@empty
\bibitem [{\citenamefont {Loss}\ and\ \citenamefont
  {DiVincenzo}(1998)}]{Loss1998}%
  \BibitemOpen
  \bibfield  {author} {\bibinfo {author} {\bibfnamefont {D.}~\bibnamefont
  {Loss}}\ and\ \bibinfo {author} {\bibfnamefont {D.~P.}\ \bibnamefont
  {DiVincenzo}},\ }\href {\doibase 10.1103/PhysRevA.57.120} {\bibfield
  {journal} {\bibinfo  {journal} {Physical Review A}\ }\textbf {\bibinfo
  {volume} {57}},\ \bibinfo {pages} {120} (\bibinfo {year} {1998})}\BibitemShut
  {NoStop}%
\bibitem [{\citenamefont {Hanson}\ \emph {et~al.}(2007)\citenamefont {Hanson},
  \citenamefont {Kouwenhoven}, \citenamefont {Petta}, \citenamefont {Tarucha},\
  and\ \citenamefont {Vandersypen}}]{Hanson2007}%
  \BibitemOpen
  \bibfield  {author} {\bibinfo {author} {\bibfnamefont {R.}~\bibnamefont
  {Hanson}}, \bibinfo {author} {\bibfnamefont {L.~P.}\ \bibnamefont
  {Kouwenhoven}}, \bibinfo {author} {\bibfnamefont {J.~R.}\ \bibnamefont
  {Petta}}, \bibinfo {author} {\bibfnamefont {S.}~\bibnamefont {Tarucha}}, \
  and\ \bibinfo {author} {\bibfnamefont {L.~M.~K.}\ \bibnamefont
  {Vandersypen}},\ }\href {\doibase 10.1103/RevModPhys.79.1217} {\bibfield
  {journal} {\bibinfo  {journal} {Reviews of Modern Physics}\ }\textbf
  {\bibinfo {volume} {79}},\ \bibinfo {pages} {1217} (\bibinfo {year}
  {2007})}\BibitemShut {NoStop}%
\bibitem [{\citenamefont {Zwanenburg}\ \emph {et~al.}(2013)\citenamefont
  {Zwanenburg}, \citenamefont {Dzurak}, \citenamefont {Morello}, \citenamefont
  {Simmons}, \citenamefont {Hollenberg}, \citenamefont {Klimeck}, \citenamefont
  {Rogge}, \citenamefont {Coppersmith},\ and\ \citenamefont
  {Eriksson}}]{Zwanenburg2013}%
  \BibitemOpen
  \bibfield  {author} {\bibinfo {author} {\bibfnamefont {F.~A.}\ \bibnamefont
  {Zwanenburg}}, \bibinfo {author} {\bibfnamefont {A.~S.}\ \bibnamefont
  {Dzurak}}, \bibinfo {author} {\bibfnamefont {A.}~\bibnamefont {Morello}},
  \bibinfo {author} {\bibfnamefont {M.~Y.}\ \bibnamefont {Simmons}}, \bibinfo
  {author} {\bibfnamefont {L.~C.~L.}\ \bibnamefont {Hollenberg}}, \bibinfo
  {author} {\bibfnamefont {G.}~\bibnamefont {Klimeck}}, \bibinfo {author}
  {\bibfnamefont {S.}~\bibnamefont {Rogge}}, \bibinfo {author} {\bibfnamefont
  {S.~N.}\ \bibnamefont {Coppersmith}}, \ and\ \bibinfo {author} {\bibfnamefont
  {M.~A.}\ \bibnamefont {Eriksson}},\ }\href {\doibase
  10.1103/RevModPhys.85.961} {\bibfield  {journal} {\bibinfo  {journal}
  {Reviews of Modern Physics}\ }\textbf {\bibinfo {volume} {85}},\ \bibinfo
  {pages} {961} (\bibinfo {year} {2013})}\BibitemShut {NoStop}%
\bibitem [{\citenamefont {Bluhm}\ \emph {et~al.}(2011)\citenamefont {Bluhm},
  \citenamefont {Foletti}, \citenamefont {Neder}, \citenamefont {Rudner},
  \citenamefont {Mahalu}, \citenamefont {Umansky},\ and\ \citenamefont
  {Yacoby}}]{Bluhm2010}%
  \BibitemOpen
  \bibfield  {author} {\bibinfo {author} {\bibfnamefont {H.}~\bibnamefont
  {Bluhm}}, \bibinfo {author} {\bibfnamefont {S.}~\bibnamefont {Foletti}},
  \bibinfo {author} {\bibfnamefont {I.}~\bibnamefont {Neder}}, \bibinfo
  {author} {\bibfnamefont {M.}~\bibnamefont {Rudner}}, \bibinfo {author}
  {\bibfnamefont {D.}~\bibnamefont {Mahalu}}, \bibinfo {author} {\bibfnamefont
  {V.}~\bibnamefont {Umansky}}, \ and\ \bibinfo {author} {\bibfnamefont
  {A.}~\bibnamefont {Yacoby}},\ }\href {\doibase 10.1038/nphys1856} {\bibfield
  {journal} {\bibinfo  {journal} {Nature Physics}\ }\textbf {\bibinfo {volume}
  {7}},\ \bibinfo {pages} {109} (\bibinfo {year} {2011})}\BibitemShut {NoStop}%
\bibitem [{\citenamefont {Maune}\ \emph {et~al.}(2012)\citenamefont {Maune},
  \citenamefont {Borselli}, \citenamefont {Huang}, \citenamefont {Ladd},
  \citenamefont {Deelman}, \citenamefont {Holabird}, \citenamefont {Kiselev},
  \citenamefont {Alvarado-Rodriguez}, \citenamefont {Ross}, \citenamefont
  {Schmitz}, \citenamefont {Sokolich}, \citenamefont {Watson}, \citenamefont
  {Gyure},\ and\ \citenamefont {Hunter}}]{Maune2012}%
  \BibitemOpen
  \bibfield  {author} {\bibinfo {author} {\bibfnamefont {B.~M.}\ \bibnamefont
  {Maune}}, \bibinfo {author} {\bibfnamefont {M.~G.}\ \bibnamefont {Borselli}},
  \bibinfo {author} {\bibfnamefont {B.}~\bibnamefont {Huang}}, \bibinfo
  {author} {\bibfnamefont {T.~D.}\ \bibnamefont {Ladd}}, \bibinfo {author}
  {\bibfnamefont {P.~W.}\ \bibnamefont {Deelman}}, \bibinfo {author}
  {\bibfnamefont {K.~S.}\ \bibnamefont {Holabird}}, \bibinfo {author}
  {\bibfnamefont {A.~A.}\ \bibnamefont {Kiselev}}, \bibinfo {author}
  {\bibfnamefont {I.}~\bibnamefont {Alvarado-Rodriguez}}, \bibinfo {author}
  {\bibfnamefont {R.~S.}\ \bibnamefont {Ross}}, \bibinfo {author}
  {\bibfnamefont {a.~E.}\ \bibnamefont {Schmitz}}, \bibinfo {author}
  {\bibfnamefont {M.}~\bibnamefont {Sokolich}}, \bibinfo {author}
  {\bibfnamefont {C.~a.}\ \bibnamefont {Watson}}, \bibinfo {author}
  {\bibfnamefont {M.~F.}\ \bibnamefont {Gyure}}, \ and\ \bibinfo {author}
  {\bibfnamefont {a.~T.}\ \bibnamefont {Hunter}},\ }\href {\doibase
  10.1038/nature10707} {\bibfield  {journal} {\bibinfo  {journal} {Nature}\
  }\textbf {\bibinfo {volume} {481}},\ \bibinfo {pages} {344} (\bibinfo {year}
  {2012})}\BibitemShut {NoStop}%
\bibitem [{\citenamefont {Veldhorst}\ \emph {et~al.}(2014)\citenamefont
  {Veldhorst}, \citenamefont {Hwang}, \citenamefont {Yang}, \citenamefont
  {Leenstra}, \citenamefont {de~Ronde}, \citenamefont {Dehollain},
  \citenamefont {Muhonen}, \citenamefont {Hudson}, \citenamefont {Itoh},
  \citenamefont {Morello},\ and\ \citenamefont {Dzurak}}]{Veldhorst2014}%
  \BibitemOpen
  \bibfield  {author} {\bibinfo {author} {\bibfnamefont {M.}~\bibnamefont
  {Veldhorst}}, \bibinfo {author} {\bibfnamefont {J.~C.~C.}\ \bibnamefont
  {Hwang}}, \bibinfo {author} {\bibfnamefont {C.~H.}\ \bibnamefont {Yang}},
  \bibinfo {author} {\bibfnamefont {A.~W.}\ \bibnamefont {Leenstra}}, \bibinfo
  {author} {\bibfnamefont {B.}~\bibnamefont {de~Ronde}}, \bibinfo {author}
  {\bibfnamefont {J.~P.}\ \bibnamefont {Dehollain}}, \bibinfo {author}
  {\bibfnamefont {J.~T.}\ \bibnamefont {Muhonen}}, \bibinfo {author}
  {\bibfnamefont {F.~E.}\ \bibnamefont {Hudson}}, \bibinfo {author}
  {\bibfnamefont {K.~M.}\ \bibnamefont {Itoh}}, \bibinfo {author}
  {\bibfnamefont {A.}~\bibnamefont {Morello}}, \ and\ \bibinfo {author}
  {\bibfnamefont {A.~S.}\ \bibnamefont {Dzurak}},\ }\href {\doibase
  10.1038/nnano.2014.216} {\bibfield  {journal} {\bibinfo  {journal} {Nature
  Nanotechnology}\ }\textbf {\bibinfo {volume} {9}},\ \bibinfo {pages} {981}
  (\bibinfo {year} {2014})}\BibitemShut {NoStop}%
\bibitem [{\citenamefont {Malinowski}\ \emph {et~al.}(2016)\citenamefont
  {Malinowski}, \citenamefont {Martins}, \citenamefont {Nissen}, \citenamefont
  {Barnes}, \citenamefont {Cywi{\'{n}}ski}, \citenamefont {Rudner},
  \citenamefont {Fallahi}, \citenamefont {Gardner}, \citenamefont {Manfra},
  \citenamefont {Marcus},\ and\ \citenamefont {Kuemmeth}}]{Malinowski2016}%
  \BibitemOpen
  \bibfield  {author} {\bibinfo {author} {\bibfnamefont {F.~K.}\ \bibnamefont
  {Malinowski}}, \bibinfo {author} {\bibfnamefont {F.}~\bibnamefont {Martins}},
  \bibinfo {author} {\bibfnamefont {P.~D.}\ \bibnamefont {Nissen}}, \bibinfo
  {author} {\bibfnamefont {E.}~\bibnamefont {Barnes}}, \bibinfo {author}
  {\bibfnamefont {{\L}.}~\bibnamefont {Cywi{\'{n}}ski}}, \bibinfo {author}
  {\bibfnamefont {M.~S.}\ \bibnamefont {Rudner}}, \bibinfo {author}
  {\bibfnamefont {S.}~\bibnamefont {Fallahi}}, \bibinfo {author} {\bibfnamefont
  {G.~C.}\ \bibnamefont {Gardner}}, \bibinfo {author} {\bibfnamefont {M.~J.}\
  \bibnamefont {Manfra}}, \bibinfo {author} {\bibfnamefont {C.~M.}\
  \bibnamefont {Marcus}}, \ and\ \bibinfo {author} {\bibfnamefont
  {F.}~\bibnamefont {Kuemmeth}},\ }\href {\doibase 10.1038/nnano.2016.170}
  {\bibfield  {journal} {\bibinfo  {journal} {Nature Nanotechnology}\ }\textbf
  {\bibinfo {volume} {12}},\ \bibinfo {pages} {16} (\bibinfo {year}
  {2016})}\BibitemShut {NoStop}%
\bibitem [{\citenamefont {Petta}\ \emph {et~al.}(2005)\citenamefont {Petta},
  \citenamefont {Johnson}, \citenamefont {Taylor}, \citenamefont {Laird},
  \citenamefont {Yacoby}, \citenamefont {Lukin}, \citenamefont {Marcus},
  \citenamefont {Hanson},\ and\ \citenamefont {Gossard}}]{Petta2005}%
  \BibitemOpen
  \bibfield  {author} {\bibinfo {author} {\bibfnamefont {J.~R.}\ \bibnamefont
  {Petta}}, \bibinfo {author} {\bibfnamefont {A.~C.}\ \bibnamefont {Johnson}},
  \bibinfo {author} {\bibfnamefont {J.~M.}\ \bibnamefont {Taylor}}, \bibinfo
  {author} {\bibfnamefont {E.~A.}\ \bibnamefont {Laird}}, \bibinfo {author}
  {\bibfnamefont {A.}~\bibnamefont {Yacoby}}, \bibinfo {author} {\bibfnamefont
  {M.~D.}\ \bibnamefont {Lukin}}, \bibinfo {author} {\bibfnamefont {C.~M.}\
  \bibnamefont {Marcus}}, \bibinfo {author} {\bibfnamefont {M.~P.}\
  \bibnamefont {Hanson}}, \ and\ \bibinfo {author} {\bibfnamefont {A.~C.}\
  \bibnamefont {Gossard}},\ }\href {\doibase 10.1126/science.1116955}
  {\bibfield  {journal} {\bibinfo  {journal} {Science}\ }\textbf {\bibinfo
  {volume} {309}},\ \bibinfo {pages} {2180} (\bibinfo {year}
  {2005})}\BibitemShut {NoStop}%
\bibitem [{\citenamefont {Foletti}\ \emph {et~al.}(2009)\citenamefont
  {Foletti}, \citenamefont {Bluhm}, \citenamefont {Mahalu}, \citenamefont
  {Umansky},\ and\ \citenamefont {Yacoby}}]{Foletti2009}%
  \BibitemOpen
  \bibfield  {author} {\bibinfo {author} {\bibfnamefont {S.}~\bibnamefont
  {Foletti}}, \bibinfo {author} {\bibfnamefont {H.}~\bibnamefont {Bluhm}},
  \bibinfo {author} {\bibfnamefont {D.}~\bibnamefont {Mahalu}}, \bibinfo
  {author} {\bibfnamefont {V.}~\bibnamefont {Umansky}}, \ and\ \bibinfo
  {author} {\bibfnamefont {A.}~\bibnamefont {Yacoby}},\ }\href {\doibase
  10.1038/nphys1424} {\bibfield  {journal} {\bibinfo  {journal} {Nature
  Physics}\ }\textbf {\bibinfo {volume} {5}},\ \bibinfo {pages} {903} (\bibinfo
  {year} {2009})}\BibitemShut {NoStop}%
\bibitem [{\citenamefont {Medford}\ \emph
  {et~al.}(2013{\natexlab{a}})\citenamefont {Medford}, \citenamefont {Beil},
  \citenamefont {Taylor}, \citenamefont {Bartlett}, \citenamefont {Doherty},
  \citenamefont {Rashba}, \citenamefont {DiVincenzo}, \citenamefont {Lu},
  \citenamefont {Gossard},\ and\ \citenamefont {Marcus}}]{Medford2013a}%
  \BibitemOpen
  \bibfield  {author} {\bibinfo {author} {\bibfnamefont {J.}~\bibnamefont
  {Medford}}, \bibinfo {author} {\bibfnamefont {J.}~\bibnamefont {Beil}},
  \bibinfo {author} {\bibfnamefont {J.~M.}\ \bibnamefont {Taylor}}, \bibinfo
  {author} {\bibfnamefont {S.~D.}\ \bibnamefont {Bartlett}}, \bibinfo {author}
  {\bibfnamefont {A.~C.}\ \bibnamefont {Doherty}}, \bibinfo {author}
  {\bibfnamefont {E.~I.}\ \bibnamefont {Rashba}}, \bibinfo {author}
  {\bibfnamefont {D.~P.}\ \bibnamefont {DiVincenzo}}, \bibinfo {author}
  {\bibfnamefont {H.}~\bibnamefont {Lu}}, \bibinfo {author} {\bibfnamefont
  {A.~C.}\ \bibnamefont {Gossard}}, \ and\ \bibinfo {author} {\bibfnamefont
  {C.~M.}\ \bibnamefont {Marcus}},\ }\href {\doibase 10.1038/nnano.2013.168}
  {\bibfield  {journal} {\bibinfo  {journal} {Nature Nanotechnology}\ }\textbf
  {\bibinfo {volume} {8}},\ \bibinfo {pages} {654} (\bibinfo {year}
  {2013}{\natexlab{a}})}\BibitemShut {NoStop}%
\bibitem [{\citenamefont {Takeda}\ \emph {et~al.}(2016)\citenamefont {Takeda},
  \citenamefont {Kamioka}, \citenamefont {Otsuka}, \citenamefont {Yoneda},
  \citenamefont {Nakajima}, \citenamefont {Delbecq}, \citenamefont {Amaha},
  \citenamefont {Allison}, \citenamefont {Kodera}, \citenamefont {Oda},\ and\
  \citenamefont {Tarucha}}]{Takeda2016a}%
  \BibitemOpen
  \bibfield  {author} {\bibinfo {author} {\bibfnamefont {K.}~\bibnamefont
  {Takeda}}, \bibinfo {author} {\bibfnamefont {J.}~\bibnamefont {Kamioka}},
  \bibinfo {author} {\bibfnamefont {T.}~\bibnamefont {Otsuka}}, \bibinfo
  {author} {\bibfnamefont {J.}~\bibnamefont {Yoneda}}, \bibinfo {author}
  {\bibfnamefont {T.}~\bibnamefont {Nakajima}}, \bibinfo {author}
  {\bibfnamefont {M.~R.}\ \bibnamefont {Delbecq}}, \bibinfo {author}
  {\bibfnamefont {S.}~\bibnamefont {Amaha}}, \bibinfo {author} {\bibfnamefont
  {G.}~\bibnamefont {Allison}}, \bibinfo {author} {\bibfnamefont
  {T.}~\bibnamefont {Kodera}}, \bibinfo {author} {\bibfnamefont
  {S.}~\bibnamefont {Oda}}, \ and\ \bibinfo {author} {\bibfnamefont
  {S.}~\bibnamefont {Tarucha}},\ }\href {\doibase 10.1126/sciadv.1600694}
  {\bibfield  {journal} {\bibinfo  {journal} {Science Advances}\ }\textbf
  {\bibinfo {volume} {2}},\ \bibinfo {pages} {e1600694} (\bibinfo {year}
  {2016})}\BibitemShut {NoStop}%
\bibitem [{\citenamefont {Kawakami}\ \emph {et~al.}(2016)\citenamefont
  {Kawakami}, \citenamefont {Jullien}, \citenamefont {Scarlino}, \citenamefont
  {Ward}, \citenamefont {Savage}, \citenamefont {Lagally}, \citenamefont
  {Dobrovitski}, \citenamefont {Friesen}, \citenamefont {Coppersmith},
  \citenamefont {Eriksson},\ and\ \citenamefont {Vandersypen}}]{Kawakami2016}%
  \BibitemOpen
  \bibfield  {author} {\bibinfo {author} {\bibfnamefont {E.}~\bibnamefont
  {Kawakami}}, \bibinfo {author} {\bibfnamefont {T.}~\bibnamefont {Jullien}},
  \bibinfo {author} {\bibfnamefont {P.}~\bibnamefont {Scarlino}}, \bibinfo
  {author} {\bibfnamefont {D.~R.}\ \bibnamefont {Ward}}, \bibinfo {author}
  {\bibfnamefont {D.~E.}\ \bibnamefont {Savage}}, \bibinfo {author}
  {\bibfnamefont {M.~G.}\ \bibnamefont {Lagally}}, \bibinfo {author}
  {\bibfnamefont {V.~V.}\ \bibnamefont {Dobrovitski}}, \bibinfo {author}
  {\bibfnamefont {M.}~\bibnamefont {Friesen}}, \bibinfo {author} {\bibfnamefont
  {S.~N.}\ \bibnamefont {Coppersmith}}, \bibinfo {author} {\bibfnamefont
  {M.~A.}\ \bibnamefont {Eriksson}}, \ and\ \bibinfo {author} {\bibfnamefont
  {L.~M.~K.}\ \bibnamefont {Vandersypen}},\ }\href {\doibase
  10.1073/pnas.1603251113} {\bibfield  {journal} {\bibinfo  {journal}
  {Proceedings of the National Academy of Sciences}\ }\textbf {\bibinfo
  {volume} {113}},\ \bibinfo {pages} {11738} (\bibinfo {year}
  {2016})}\BibitemShut {NoStop}%
\bibitem [{\citenamefont {Nowack}\ \emph {et~al.}(2011)\citenamefont {Nowack},
  \citenamefont {Shafiei}, \citenamefont {Laforest}, \citenamefont
  {Prawiroatmodjo}, \citenamefont {Schreiber}, \citenamefont {Reichl},
  \citenamefont {Wegscheider},\ and\ \citenamefont {Vandersypen}}]{Nowack2011}%
  \BibitemOpen
  \bibfield  {author} {\bibinfo {author} {\bibfnamefont {K.~C.}\ \bibnamefont
  {Nowack}}, \bibinfo {author} {\bibfnamefont {M.}~\bibnamefont {Shafiei}},
  \bibinfo {author} {\bibfnamefont {M.}~\bibnamefont {Laforest}}, \bibinfo
  {author} {\bibfnamefont {G.~E. D.~K.}\ \bibnamefont {Prawiroatmodjo}},
  \bibinfo {author} {\bibfnamefont {L.~R.}\ \bibnamefont {Schreiber}}, \bibinfo
  {author} {\bibfnamefont {C.}~\bibnamefont {Reichl}}, \bibinfo {author}
  {\bibfnamefont {W.}~\bibnamefont {Wegscheider}}, \ and\ \bibinfo {author}
  {\bibfnamefont {L.~M.~K.}\ \bibnamefont {Vandersypen}},\ }\href {\doibase
  10.1126/science.1209524} {\bibfield  {journal} {\bibinfo  {journal}
  {Science}\ }\textbf {\bibinfo {volume} {333}},\ \bibinfo {pages} {1269}
  (\bibinfo {year} {2011})}\BibitemShut {NoStop}%
\bibitem [{\citenamefont {Veldhorst}\ \emph {et~al.}(2015)\citenamefont
  {Veldhorst}, \citenamefont {Yang}, \citenamefont {Hwang}, \citenamefont
  {Huang}, \citenamefont {Dehollain}, \citenamefont {Muhonen}, \citenamefont
  {Simmons}, \citenamefont {Laucht}, \citenamefont {Hudson}, \citenamefont
  {Itoh}, \citenamefont {Morello},\ and\ \citenamefont
  {Dzurak}}]{Veldhorst2015a}%
  \BibitemOpen
  \bibfield  {author} {\bibinfo {author} {\bibfnamefont {M.}~\bibnamefont
  {Veldhorst}}, \bibinfo {author} {\bibfnamefont {C.~H.}\ \bibnamefont {Yang}},
  \bibinfo {author} {\bibfnamefont {J.~C.~C.}\ \bibnamefont {Hwang}}, \bibinfo
  {author} {\bibfnamefont {W.}~\bibnamefont {Huang}}, \bibinfo {author}
  {\bibfnamefont {J.~P.}\ \bibnamefont {Dehollain}}, \bibinfo {author}
  {\bibfnamefont {J.~T.}\ \bibnamefont {Muhonen}}, \bibinfo {author}
  {\bibfnamefont {S.}~\bibnamefont {Simmons}}, \bibinfo {author} {\bibfnamefont
  {A.}~\bibnamefont {Laucht}}, \bibinfo {author} {\bibfnamefont {F.~E.}\
  \bibnamefont {Hudson}}, \bibinfo {author} {\bibfnamefont {K.~M.}\
  \bibnamefont {Itoh}}, \bibinfo {author} {\bibfnamefont {A.}~\bibnamefont
  {Morello}}, \ and\ \bibinfo {author} {\bibfnamefont {A.~S.}\ \bibnamefont
  {Dzurak}},\ }\href {\doibase 10.1038/nature15263} {\bibfield  {journal}
  {\bibinfo  {journal} {Nature}\ }\textbf {\bibinfo {volume} {526}},\ \bibinfo
  {pages} {410} (\bibinfo {year} {2015})}\BibitemShut {NoStop}%
\bibitem [{\citenamefont {Zajac}\ \emph {et~al.}(2017)\citenamefont {Zajac},
  \citenamefont {Sigillito}, \citenamefont {Russ}, \citenamefont {Borjans},
  \citenamefont {Taylor}, \citenamefont {Burkard},\ and\ \citenamefont
  {Petta}}]{Zajac2017}%
  \BibitemOpen
  \bibfield  {author} {\bibinfo {author} {\bibfnamefont {D.~M.}\ \bibnamefont
  {Zajac}}, \bibinfo {author} {\bibfnamefont {A.~J.}\ \bibnamefont
  {Sigillito}}, \bibinfo {author} {\bibfnamefont {M.}~\bibnamefont {Russ}},
  \bibinfo {author} {\bibfnamefont {F.}~\bibnamefont {Borjans}}, \bibinfo
  {author} {\bibfnamefont {J.~M.}\ \bibnamefont {Taylor}}, \bibinfo {author}
  {\bibfnamefont {G.}~\bibnamefont {Burkard}}, \ and\ \bibinfo {author}
  {\bibfnamefont {J.~R.}\ \bibnamefont {Petta}},\ }\href
  {http://arxiv.org/abs/1708.03530} {\  (\bibinfo {year} {2017})},\ \Eprint
  {http://arxiv.org/abs/1708.03530} {arXiv:1708.03530} \BibitemShut {NoStop}%
\bibitem [{\citenamefont {Cerfontaine}\ \emph {et~al.}(2014)\citenamefont
  {Cerfontaine}, \citenamefont {Botzem}, \citenamefont {DiVincenzo},\ and\
  \citenamefont {Bluhm}}]{Cerfontaine2014}%
  \BibitemOpen
  \bibfield  {author} {\bibinfo {author} {\bibfnamefont {P.}~\bibnamefont
  {Cerfontaine}}, \bibinfo {author} {\bibfnamefont {T.}~\bibnamefont {Botzem}},
  \bibinfo {author} {\bibfnamefont {D.~P.}\ \bibnamefont {DiVincenzo}}, \ and\
  \bibinfo {author} {\bibfnamefont {H.}~\bibnamefont {Bluhm}},\ }\href
  {\doibase 10.1103/PhysRevLett.113.150501} {\bibfield  {journal} {\bibinfo
  {journal} {Physical Review Letters}\ }\textbf {\bibinfo {volume} {113}},\
  \bibinfo {pages} {150501} (\bibinfo {year} {2014})}\BibitemShut {NoStop}%
\bibitem [{\citenamefont {Wu}\ \emph {et~al.}(2014)\citenamefont {Wu},
  \citenamefont {Ward}, \citenamefont {Prance}, \citenamefont {Kim},
  \citenamefont {Gamble}, \citenamefont {Mohr}, \citenamefont {Shi},
  \citenamefont {Savage}, \citenamefont {Lagally}, \citenamefont {Friesen},
  \citenamefont {Coppersmith},\ and\ \citenamefont {Eriksson}}]{Wu2014}%
  \BibitemOpen
  \bibfield  {author} {\bibinfo {author} {\bibfnamefont {X.}~\bibnamefont
  {Wu}}, \bibinfo {author} {\bibfnamefont {D.~R.}\ \bibnamefont {Ward}},
  \bibinfo {author} {\bibfnamefont {J.~R.}\ \bibnamefont {Prance}}, \bibinfo
  {author} {\bibfnamefont {D.}~\bibnamefont {Kim}}, \bibinfo {author}
  {\bibfnamefont {J.~K.}\ \bibnamefont {Gamble}}, \bibinfo {author}
  {\bibfnamefont {R.~T.}\ \bibnamefont {Mohr}}, \bibinfo {author}
  {\bibfnamefont {Z.}~\bibnamefont {Shi}}, \bibinfo {author} {\bibfnamefont
  {D.~E.}\ \bibnamefont {Savage}}, \bibinfo {author} {\bibfnamefont {M.~G.}\
  \bibnamefont {Lagally}}, \bibinfo {author} {\bibfnamefont {M.}~\bibnamefont
  {Friesen}}, \bibinfo {author} {\bibfnamefont {S.~N.}\ \bibnamefont
  {Coppersmith}}, \ and\ \bibinfo {author} {\bibfnamefont {M.~A.}\ \bibnamefont
  {Eriksson}},\ }\href {\doibase 10.1073/pnas.1412230111} {\bibfield  {journal}
  {\bibinfo  {journal} {Proceedings of the National Academy of Sciences}\
  }\textbf {\bibinfo {volume} {111}},\ \bibinfo {pages} {11938} (\bibinfo
  {year} {2014})}\BibitemShut {NoStop}%
\bibitem [{\citenamefont {Shulman}\ \emph {et~al.}(2012)\citenamefont
  {Shulman}, \citenamefont {Dial}, \citenamefont {Harvey}, \citenamefont
  {Bluhm}, \citenamefont {Umansky},\ and\ \citenamefont
  {Yacoby}}]{Shulman2012}%
  \BibitemOpen
  \bibfield  {author} {\bibinfo {author} {\bibfnamefont {M.~D.}\ \bibnamefont
  {Shulman}}, \bibinfo {author} {\bibfnamefont {O.~E.}\ \bibnamefont {Dial}},
  \bibinfo {author} {\bibfnamefont {S.~P.}\ \bibnamefont {Harvey}}, \bibinfo
  {author} {\bibfnamefont {H.}~\bibnamefont {Bluhm}}, \bibinfo {author}
  {\bibfnamefont {V.}~\bibnamefont {Umansky}}, \ and\ \bibinfo {author}
  {\bibfnamefont {A.}~\bibnamefont {Yacoby}},\ }\href {\doibase
  10.1126/science.1217692} {\bibfield  {journal} {\bibinfo  {journal}
  {Science}\ }\textbf {\bibinfo {volume} {336}},\ \bibinfo {pages} {202}
  (\bibinfo {year} {2012})}\BibitemShut {NoStop}%
\bibitem [{\citenamefont {Nichol}\ \emph {et~al.}(2017)\citenamefont {Nichol},
  \citenamefont {Orona}, \citenamefont {Harvey}, \citenamefont {Fallahi},
  \citenamefont {Gardner}, \citenamefont {Manfra},\ and\ \citenamefont
  {Yacoby}}]{Nichol2016}%
  \BibitemOpen
  \bibfield  {author} {\bibinfo {author} {\bibfnamefont {J.~M.}\ \bibnamefont
  {Nichol}}, \bibinfo {author} {\bibfnamefont {L.~A.}\ \bibnamefont {Orona}},
  \bibinfo {author} {\bibfnamefont {S.~P.}\ \bibnamefont {Harvey}}, \bibinfo
  {author} {\bibfnamefont {S.}~\bibnamefont {Fallahi}}, \bibinfo {author}
  {\bibfnamefont {G.~C.}\ \bibnamefont {Gardner}}, \bibinfo {author}
  {\bibfnamefont {M.~J.}\ \bibnamefont {Manfra}}, \ and\ \bibinfo {author}
  {\bibfnamefont {A.}~\bibnamefont {Yacoby}},\ }\href {\doibase
  10.1038/s41534-016-0003-1} {\bibfield  {journal} {\bibinfo  {journal} {npj
  Quantum Information}\ }\textbf {\bibinfo {volume} {3}},\ \bibinfo {pages} {3}
  (\bibinfo {year} {2017})}\BibitemShut {NoStop}%
\bibitem [{\citenamefont {Laird}\ \emph {et~al.}(2010)\citenamefont {Laird},
  \citenamefont {Taylor}, \citenamefont {DiVincenzo}, \citenamefont {Marcus},
  \citenamefont {Hanson},\ and\ \citenamefont {Gossard}}]{Laird2010}%
  \BibitemOpen
  \bibfield  {author} {\bibinfo {author} {\bibfnamefont {E.~A.}\ \bibnamefont
  {Laird}}, \bibinfo {author} {\bibfnamefont {J.~M.}\ \bibnamefont {Taylor}},
  \bibinfo {author} {\bibfnamefont {D.~P.}\ \bibnamefont {DiVincenzo}},
  \bibinfo {author} {\bibfnamefont {C.~M.}\ \bibnamefont {Marcus}}, \bibinfo
  {author} {\bibfnamefont {M.~P.}\ \bibnamefont {Hanson}}, \ and\ \bibinfo
  {author} {\bibfnamefont {A.~C.}\ \bibnamefont {Gossard}},\ }\href {\doibase
  10.1103/PhysRevB.82.075403} {\bibfield  {journal} {\bibinfo  {journal}
  {Physical Review B}\ }\textbf {\bibinfo {volume} {82}},\ \bibinfo {pages}
  {075403} (\bibinfo {year} {2010})}\BibitemShut {NoStop}%
\bibitem [{\citenamefont {Medford}\ \emph
  {et~al.}(2013{\natexlab{b}})\citenamefont {Medford}, \citenamefont {Beil},
  \citenamefont {Taylor}, \citenamefont {Rashba}, \citenamefont {Lu},
  \citenamefont {Gossard},\ and\ \citenamefont {Marcus}}]{Medford2013}%
  \BibitemOpen
  \bibfield  {author} {\bibinfo {author} {\bibfnamefont {J.}~\bibnamefont
  {Medford}}, \bibinfo {author} {\bibfnamefont {J.}~\bibnamefont {Beil}},
  \bibinfo {author} {\bibfnamefont {J.~M.}\ \bibnamefont {Taylor}}, \bibinfo
  {author} {\bibfnamefont {E.~I.}\ \bibnamefont {Rashba}}, \bibinfo {author}
  {\bibfnamefont {H.}~\bibnamefont {Lu}}, \bibinfo {author} {\bibfnamefont
  {A.~C.}\ \bibnamefont {Gossard}}, \ and\ \bibinfo {author} {\bibfnamefont
  {C.~M.}\ \bibnamefont {Marcus}},\ }\href {\doibase
  10.1103/PhysRevLett.111.050501} {\bibfield  {journal} {\bibinfo  {journal}
  {Physical Review Letters}\ }\textbf {\bibinfo {volume} {111}},\ \bibinfo
  {pages} {050501} (\bibinfo {year} {2013}{\natexlab{b}})}\BibitemShut
  {NoStop}%
\bibitem [{\citenamefont {Eng}\ \emph {et~al.}(2015)\citenamefont {Eng},
  \citenamefont {Ladd}, \citenamefont {Smith}, \citenamefont {Borselli},
  \citenamefont {Kiselev}, \citenamefont {Fong}, \citenamefont {Holabird},
  \citenamefont {Hazard}, \citenamefont {Huang}, \citenamefont {Deelman},
  \citenamefont {Milosavljevic}, \citenamefont {Schmitz}, \citenamefont {Ross},
  \citenamefont {Gyure},\ and\ \citenamefont {Hunter}}]{Eng2015}%
  \BibitemOpen
  \bibfield  {author} {\bibinfo {author} {\bibfnamefont {K.}~\bibnamefont
  {Eng}}, \bibinfo {author} {\bibfnamefont {T.~D.}\ \bibnamefont {Ladd}},
  \bibinfo {author} {\bibfnamefont {A.}~\bibnamefont {Smith}}, \bibinfo
  {author} {\bibfnamefont {M.~G.}\ \bibnamefont {Borselli}}, \bibinfo {author}
  {\bibfnamefont {A.~A.}\ \bibnamefont {Kiselev}}, \bibinfo {author}
  {\bibfnamefont {B.~H.}\ \bibnamefont {Fong}}, \bibinfo {author}
  {\bibfnamefont {K.~S.}\ \bibnamefont {Holabird}}, \bibinfo {author}
  {\bibfnamefont {T.~M.}\ \bibnamefont {Hazard}}, \bibinfo {author}
  {\bibfnamefont {B.}~\bibnamefont {Huang}}, \bibinfo {author} {\bibfnamefont
  {P.~W.}\ \bibnamefont {Deelman}}, \bibinfo {author} {\bibfnamefont
  {I.}~\bibnamefont {Milosavljevic}}, \bibinfo {author} {\bibfnamefont {A.~E.}\
  \bibnamefont {Schmitz}}, \bibinfo {author} {\bibfnamefont {R.~S.}\
  \bibnamefont {Ross}}, \bibinfo {author} {\bibfnamefont {M.~F.}\ \bibnamefont
  {Gyure}}, \ and\ \bibinfo {author} {\bibfnamefont {A.~T.}\ \bibnamefont
  {Hunter}},\ }\href {\doibase 10.1126/sciadv.1500214} {\bibfield  {journal}
  {\bibinfo  {journal} {Science Advances}\ }\textbf {\bibinfo {volume} {1}},\
  \bibinfo {pages} {e1500214} (\bibinfo {year} {2015})}\BibitemShut {NoStop}%
\bibitem [{\citenamefont {Shi}\ \emph {et~al.}(2012)\citenamefont {Shi},
  \citenamefont {Simmons}, \citenamefont {Prance}, \citenamefont {Gamble},
  \citenamefont {Koh}, \citenamefont {Shim}, \citenamefont {Hu}, \citenamefont
  {Savage}, \citenamefont {Lagally}, \citenamefont {Eriksson}, \citenamefont
  {Friesen},\ and\ \citenamefont {Coppersmith}}]{Shi2012}%
  \BibitemOpen
  \bibfield  {author} {\bibinfo {author} {\bibfnamefont {Z.}~\bibnamefont
  {Shi}}, \bibinfo {author} {\bibfnamefont {C.~B.}\ \bibnamefont {Simmons}},
  \bibinfo {author} {\bibfnamefont {J.~R.}\ \bibnamefont {Prance}}, \bibinfo
  {author} {\bibfnamefont {J.~K.}\ \bibnamefont {Gamble}}, \bibinfo {author}
  {\bibfnamefont {T.~S.}\ \bibnamefont {Koh}}, \bibinfo {author} {\bibfnamefont
  {Y.-P.}\ \bibnamefont {Shim}}, \bibinfo {author} {\bibfnamefont
  {X.}~\bibnamefont {Hu}}, \bibinfo {author} {\bibfnamefont {D.~E.}\
  \bibnamefont {Savage}}, \bibinfo {author} {\bibfnamefont {M.~G.}\
  \bibnamefont {Lagally}}, \bibinfo {author} {\bibfnamefont {M.~A.}\
  \bibnamefont {Eriksson}}, \bibinfo {author} {\bibfnamefont {M.}~\bibnamefont
  {Friesen}}, \ and\ \bibinfo {author} {\bibfnamefont {S.~N.}\ \bibnamefont
  {Coppersmith}},\ }\href {\doibase 10.1103/PhysRevLett.108.140503} {\bibfield
  {journal} {\bibinfo  {journal} {Physical Review Letters}\ }\textbf {\bibinfo
  {volume} {108}},\ \bibinfo {pages} {140503} (\bibinfo {year}
  {2012})}\BibitemShut {NoStop}%
\bibitem [{\citenamefont {Koh}\ \emph {et~al.}(2012)\citenamefont {Koh},
  \citenamefont {Gamble}, \citenamefont {Friesen}, \citenamefont {Eriksson},\
  and\ \citenamefont {Coppersmith}}]{Koh2012}%
  \BibitemOpen
  \bibfield  {author} {\bibinfo {author} {\bibfnamefont {T.~S.}\ \bibnamefont
  {Koh}}, \bibinfo {author} {\bibfnamefont {J.~K.}\ \bibnamefont {Gamble}},
  \bibinfo {author} {\bibfnamefont {M.}~\bibnamefont {Friesen}}, \bibinfo
  {author} {\bibfnamefont {M.~A.}\ \bibnamefont {Eriksson}}, \ and\ \bibinfo
  {author} {\bibfnamefont {S.~N.}\ \bibnamefont {Coppersmith}},\ }\href
  {\doibase 10.1103/PhysRevLett.109.250503} {\bibfield  {journal} {\bibinfo
  {journal} {Physical Review Letters}\ }\textbf {\bibinfo {volume} {109}},\
  \bibinfo {pages} {250503} (\bibinfo {year} {2012})}\BibitemShut {NoStop}%
\bibitem [{\citenamefont {Kim}\ \emph {et~al.}(2014)\citenamefont {Kim},
  \citenamefont {Shi}, \citenamefont {Simmons}, \citenamefont {Ward},
  \citenamefont {Prance}, \citenamefont {Koh}, \citenamefont {Gamble},
  \citenamefont {Savage}, \citenamefont {Lagally}, \citenamefont {Friesen},
  \citenamefont {Coppersmith},\ and\ \citenamefont {Eriksson}}]{Kim2014}%
  \BibitemOpen
  \bibfield  {author} {\bibinfo {author} {\bibfnamefont {D.}~\bibnamefont
  {Kim}}, \bibinfo {author} {\bibfnamefont {Z.}~\bibnamefont {Shi}}, \bibinfo
  {author} {\bibfnamefont {C.~B.}\ \bibnamefont {Simmons}}, \bibinfo {author}
  {\bibfnamefont {D.~R.}\ \bibnamefont {Ward}}, \bibinfo {author}
  {\bibfnamefont {J.~R.}\ \bibnamefont {Prance}}, \bibinfo {author}
  {\bibfnamefont {T.~S.}\ \bibnamefont {Koh}}, \bibinfo {author} {\bibfnamefont
  {J.~K.}\ \bibnamefont {Gamble}}, \bibinfo {author} {\bibfnamefont {D.~E.}\
  \bibnamefont {Savage}}, \bibinfo {author} {\bibfnamefont {M.~G.}\
  \bibnamefont {Lagally}}, \bibinfo {author} {\bibfnamefont {M.}~\bibnamefont
  {Friesen}}, \bibinfo {author} {\bibfnamefont {S.~N.}\ \bibnamefont
  {Coppersmith}}, \ and\ \bibinfo {author} {\bibfnamefont {M.~A.}\ \bibnamefont
  {Eriksson}},\ }\href {\doibase 10.1038/nature13407} {\bibfield  {journal}
  {\bibinfo  {journal} {Nature}\ }\textbf {\bibinfo {volume} {511}},\ \bibinfo
  {pages} {70} (\bibinfo {year} {2014})}\BibitemShut {NoStop}%
\bibitem [{\citenamefont {Kim}\ \emph {et~al.}(2015)\citenamefont {Kim},
  \citenamefont {Ward}, \citenamefont {Simmons}, \citenamefont {Savage},
  \citenamefont {Lagally}, \citenamefont {Friesen}, \citenamefont
  {Coppersmith},\ and\ \citenamefont {Eriksson}}]{Kim2015b}%
  \BibitemOpen
  \bibfield  {author} {\bibinfo {author} {\bibfnamefont {D.}~\bibnamefont
  {Kim}}, \bibinfo {author} {\bibfnamefont {D.~R.}\ \bibnamefont {Ward}},
  \bibinfo {author} {\bibfnamefont {C.~B.}\ \bibnamefont {Simmons}}, \bibinfo
  {author} {\bibfnamefont {D.~E.}\ \bibnamefont {Savage}}, \bibinfo {author}
  {\bibfnamefont {M.~G.}\ \bibnamefont {Lagally}}, \bibinfo {author}
  {\bibfnamefont {M.}~\bibnamefont {Friesen}}, \bibinfo {author} {\bibfnamefont
  {S.~N.}\ \bibnamefont {Coppersmith}}, \ and\ \bibinfo {author} {\bibfnamefont
  {M.~A.}\ \bibnamefont {Eriksson}},\ }\href {\doibase 10.1038/npjqi.2015.4}
  {\bibfield  {journal} {\bibinfo  {journal} {npj Quantum Information}\
  }\textbf {\bibinfo {volume} {1}},\ \bibinfo {pages} {15004} (\bibinfo {year}
  {2015})}\BibitemShut {NoStop}%
\bibitem [{\citenamefont {Cao}\ \emph {et~al.}(2016)\citenamefont {Cao},
  \citenamefont {Li}, \citenamefont {Yu}, \citenamefont {Wang}, \citenamefont
  {Chen}, \citenamefont {Song}, \citenamefont {Xiao}, \citenamefont {Guo},
  \citenamefont {Jiang}, \citenamefont {Hu},\ and\ \citenamefont
  {Guo}}]{Cao2016}%
  \BibitemOpen
  \bibfield  {author} {\bibinfo {author} {\bibfnamefont {G.}~\bibnamefont
  {Cao}}, \bibinfo {author} {\bibfnamefont {H.-O.}\ \bibnamefont {Li}},
  \bibinfo {author} {\bibfnamefont {G.-D.}\ \bibnamefont {Yu}}, \bibinfo
  {author} {\bibfnamefont {B.-C.}\ \bibnamefont {Wang}}, \bibinfo {author}
  {\bibfnamefont {B.-B.}\ \bibnamefont {Chen}}, \bibinfo {author}
  {\bibfnamefont {X.-X.}\ \bibnamefont {Song}}, \bibinfo {author}
  {\bibfnamefont {M.}~\bibnamefont {Xiao}}, \bibinfo {author} {\bibfnamefont
  {G.-C.}\ \bibnamefont {Guo}}, \bibinfo {author} {\bibfnamefont {H.-W.}\
  \bibnamefont {Jiang}}, \bibinfo {author} {\bibfnamefont {X.}~\bibnamefont
  {Hu}}, \ and\ \bibinfo {author} {\bibfnamefont {G.-P.}\ \bibnamefont {Guo}},\
  }\href {\doibase 10.1103/PhysRevLett.116.086801} {\bibfield  {journal}
  {\bibinfo  {journal} {Physical Review Letters}\ }\textbf {\bibinfo {volume}
  {116}},\ \bibinfo {pages} {086801} (\bibinfo {year} {2016})}\BibitemShut
  {NoStop}%
\bibitem [{\citenamefont {Brunner}\ \emph {et~al.}(2011)\citenamefont
  {Brunner}, \citenamefont {Shin}, \citenamefont {Obata}, \citenamefont
  {Pioro-Ladri{\`{e}}re}, \citenamefont {Kubo}, \citenamefont {Yoshida},
  \citenamefont {Taniyama}, \citenamefont {Tokura},\ and\ \citenamefont
  {Tarucha}}]{Brunner2011}%
  \BibitemOpen
  \bibfield  {author} {\bibinfo {author} {\bibfnamefont {R.}~\bibnamefont
  {Brunner}}, \bibinfo {author} {\bibfnamefont {Y.-S.}\ \bibnamefont {Shin}},
  \bibinfo {author} {\bibfnamefont {T.}~\bibnamefont {Obata}}, \bibinfo
  {author} {\bibfnamefont {M.}~\bibnamefont {Pioro-Ladri{\`{e}}re}}, \bibinfo
  {author} {\bibfnamefont {T.}~\bibnamefont {Kubo}}, \bibinfo {author}
  {\bibfnamefont {K.}~\bibnamefont {Yoshida}}, \bibinfo {author} {\bibfnamefont
  {T.}~\bibnamefont {Taniyama}}, \bibinfo {author} {\bibfnamefont
  {Y.}~\bibnamefont {Tokura}}, \ and\ \bibinfo {author} {\bibfnamefont
  {S.}~\bibnamefont {Tarucha}},\ }\href {\doibase
  10.1103/PhysRevLett.107.146801} {\bibfield  {journal} {\bibinfo  {journal}
  {Physical Review Letters}\ }\textbf {\bibinfo {volume} {107}},\ \bibinfo
  {pages} {146801} (\bibinfo {year} {2011})}\BibitemShut {NoStop}%
\bibitem [{\citenamefont {Hu}\ and\ \citenamefont {{Das
  Sarma}}(2006)}]{Hu2006}%
  \BibitemOpen
  \bibfield  {author} {\bibinfo {author} {\bibfnamefont {X.}~\bibnamefont
  {Hu}}\ and\ \bibinfo {author} {\bibfnamefont {S.}~\bibnamefont {{Das
  Sarma}}},\ }\href {\doibase 10.1103/PhysRevLett.96.100501} {\bibfield
  {journal} {\bibinfo  {journal} {Physical Review Letters}\ }\textbf {\bibinfo
  {volume} {96}},\ \bibinfo {pages} {100501} (\bibinfo {year}
  {2006})}\BibitemShut {NoStop}%
\bibitem [{\citenamefont {Reed}\ \emph {et~al.}(2016)\citenamefont {Reed},
  \citenamefont {Maune}, \citenamefont {Andrews}, \citenamefont {Borselli},
  \citenamefont {Eng}, \citenamefont {Jura}, \citenamefont {Kiselev},
  \citenamefont {Ladd}, \citenamefont {Merkel}, \citenamefont {Milosavljevic},
  \citenamefont {Pritchett}, \citenamefont {Rakher}, \citenamefont {Ross},
  \citenamefont {Schmitz}, \citenamefont {Smith}, \citenamefont {Wright},
  \citenamefont {Gyure},\ and\ \citenamefont {Hunter}}]{Reed2016}%
  \BibitemOpen
  \bibfield  {author} {\bibinfo {author} {\bibfnamefont {M.~D.}\ \bibnamefont
  {Reed}}, \bibinfo {author} {\bibfnamefont {B.~M.}\ \bibnamefont {Maune}},
  \bibinfo {author} {\bibfnamefont {R.~W.}\ \bibnamefont {Andrews}}, \bibinfo
  {author} {\bibfnamefont {M.~G.}\ \bibnamefont {Borselli}}, \bibinfo {author}
  {\bibfnamefont {K.}~\bibnamefont {Eng}}, \bibinfo {author} {\bibfnamefont
  {M.~P.}\ \bibnamefont {Jura}}, \bibinfo {author} {\bibfnamefont {A.~A.}\
  \bibnamefont {Kiselev}}, \bibinfo {author} {\bibfnamefont {T.~D.}\
  \bibnamefont {Ladd}}, \bibinfo {author} {\bibfnamefont {S.~T.}\ \bibnamefont
  {Merkel}}, \bibinfo {author} {\bibfnamefont {I.}~\bibnamefont
  {Milosavljevic}}, \bibinfo {author} {\bibfnamefont {E.~J.}\ \bibnamefont
  {Pritchett}}, \bibinfo {author} {\bibfnamefont {M.~T.}\ \bibnamefont
  {Rakher}}, \bibinfo {author} {\bibfnamefont {R.~S.}\ \bibnamefont {Ross}},
  \bibinfo {author} {\bibfnamefont {A.~E.}\ \bibnamefont {Schmitz}}, \bibinfo
  {author} {\bibfnamefont {A.}~\bibnamefont {Smith}}, \bibinfo {author}
  {\bibfnamefont {J.~A.}\ \bibnamefont {Wright}}, \bibinfo {author}
  {\bibfnamefont {M.~F.}\ \bibnamefont {Gyure}}, \ and\ \bibinfo {author}
  {\bibfnamefont {A.~T.}\ \bibnamefont {Hunter}},\ }\href {\doibase
  10.1103/PhysRevLett.116.110402} {\bibfield  {journal} {\bibinfo  {journal}
  {Physical Review Letters}\ }\textbf {\bibinfo {volume} {116}},\ \bibinfo
  {pages} {110402} (\bibinfo {year} {2016})}\BibitemShut {NoStop}%
\bibitem [{\citenamefont {Martins}\ \emph {et~al.}(2016)\citenamefont
  {Martins}, \citenamefont {Malinowski}, \citenamefont {Nissen}, \citenamefont
  {Barnes}, \citenamefont {Fallahi}, \citenamefont {Gardner}, \citenamefont
  {Manfra}, \citenamefont {Marcus},\ and\ \citenamefont
  {Kuemmeth}}]{Martins2016}%
  \BibitemOpen
  \bibfield  {author} {\bibinfo {author} {\bibfnamefont {F.}~\bibnamefont
  {Martins}}, \bibinfo {author} {\bibfnamefont {F.~K.}\ \bibnamefont
  {Malinowski}}, \bibinfo {author} {\bibfnamefont {P.~D.}\ \bibnamefont
  {Nissen}}, \bibinfo {author} {\bibfnamefont {E.}~\bibnamefont {Barnes}},
  \bibinfo {author} {\bibfnamefont {S.}~\bibnamefont {Fallahi}}, \bibinfo
  {author} {\bibfnamefont {G.~C.}\ \bibnamefont {Gardner}}, \bibinfo {author}
  {\bibfnamefont {M.~J.}\ \bibnamefont {Manfra}}, \bibinfo {author}
  {\bibfnamefont {C.~M.}\ \bibnamefont {Marcus}}, \ and\ \bibinfo {author}
  {\bibfnamefont {F.}~\bibnamefont {Kuemmeth}},\ }\href {\doibase
  10.1103/PhysRevLett.116.116801} {\bibfield  {journal} {\bibinfo  {journal}
  {Physical Review Letters}\ }\textbf {\bibinfo {volume} {116}},\ \bibinfo
  {pages} {116801} (\bibinfo {year} {2016})}\BibitemShut {NoStop}%
\bibitem [{\citenamefont {Barnes}\ \emph {et~al.}(2016)\citenamefont {Barnes},
  \citenamefont {Rudner}, \citenamefont {Martins}, \citenamefont {Malinowski},
  \citenamefont {Marcus},\ and\ \citenamefont {Kuemmeth}}]{Barnes2016}%
  \BibitemOpen
  \bibfield  {author} {\bibinfo {author} {\bibfnamefont {E.}~\bibnamefont
  {Barnes}}, \bibinfo {author} {\bibfnamefont {M.~S.}\ \bibnamefont {Rudner}},
  \bibinfo {author} {\bibfnamefont {F.}~\bibnamefont {Martins}}, \bibinfo
  {author} {\bibfnamefont {F.~K.}\ \bibnamefont {Malinowski}}, \bibinfo
  {author} {\bibfnamefont {C.~M.}\ \bibnamefont {Marcus}}, \ and\ \bibinfo
  {author} {\bibfnamefont {F.}~\bibnamefont {Kuemmeth}},\ }\href {\doibase
  10.1103/PhysRevB.93.121407} {\bibfield  {journal} {\bibinfo  {journal}
  {Physical Review B}\ }\textbf {\bibinfo {volume} {93}},\ \bibinfo {pages}
  {121407} (\bibinfo {year} {2016})}\BibitemShut {NoStop}%
\bibitem [{\citenamefont {Zhang}\ \emph {et~al.}(2017)\citenamefont {Zhang},
  \citenamefont {Throckmorton}, \citenamefont {Yang}, \citenamefont {Wang},
  \citenamefont {Barnes},\ and\ \citenamefont {{Das Sarma}}}]{Zhang2017}%
  \BibitemOpen
  \bibfield  {author} {\bibinfo {author} {\bibfnamefont {C.}~\bibnamefont
  {Zhang}}, \bibinfo {author} {\bibfnamefont {R.~E.}\ \bibnamefont
  {Throckmorton}}, \bibinfo {author} {\bibfnamefont {X.-C.}\ \bibnamefont
  {Yang}}, \bibinfo {author} {\bibfnamefont {X.}~\bibnamefont {Wang}}, \bibinfo
  {author} {\bibfnamefont {E.}~\bibnamefont {Barnes}}, \ and\ \bibinfo {author}
  {\bibfnamefont {S.}~\bibnamefont {{Das Sarma}}},\ }\href {\doibase
  10.1103/PhysRevLett.118.216802} {\bibfield  {journal} {\bibinfo  {journal}
  {Physical Review Letters}\ }\textbf {\bibinfo {volume} {118}},\ \bibinfo
  {pages} {216802} (\bibinfo {year} {2017})}\BibitemShut {NoStop}%
\bibitem [{\citenamefont {Yang}\ and\ \citenamefont
  {Wang}(2017{\natexlab{a}})}]{Yang2017}%
  \BibitemOpen
  \bibfield  {author} {\bibinfo {author} {\bibfnamefont {X.-C.}\ \bibnamefont
  {Yang}}\ and\ \bibinfo {author} {\bibfnamefont {X.}~\bibnamefont {Wang}},\
  }\href {\doibase 10.1103/PhysRevA.96.012318} {\bibfield  {journal} {\bibinfo
  {journal} {Physical Review A}\ }\textbf {\bibinfo {volume} {96}},\ \bibinfo
  {pages} {012318} (\bibinfo {year} {2017}{\natexlab{a}})}\BibitemShut
  {NoStop}%
\bibitem [{\citenamefont {Yang}\ and\ \citenamefont
  {Wang}(2017{\natexlab{b}})}]{Yang2017a}%
  \BibitemOpen
  \bibfield  {author} {\bibinfo {author} {\bibfnamefont {X.-c.}\ \bibnamefont
  {Yang}}\ and\ \bibinfo {author} {\bibfnamefont {X.}~\bibnamefont {Wang}},\
  }\href {http://arxiv.org/abs/1707.07929} {\ \textbf {\bibinfo {volume} {1}},\
  \bibinfo {pages} {1} (\bibinfo {year} {2017}{\natexlab{b}})},\ \Eprint
  {http://arxiv.org/abs/1707.07929} {arXiv:1707.07929} \BibitemShut {NoStop}%
\bibitem [{\citenamefont {Shim}\ and\ \citenamefont {Tahan}(2017)}]{Shim2017}%
  \BibitemOpen
  \bibfield  {author} {\bibinfo {author} {\bibfnamefont {Y.-P.}\ \bibnamefont
  {Shim}}\ and\ \bibinfo {author} {\bibfnamefont {C.}~\bibnamefont {Tahan}},\
  }\href {http://arxiv.org/abs/1711.00595} {\ ,\ \bibinfo {pages} {1} (\bibinfo
  {year} {2017})},\ \Eprint {http://arxiv.org/abs/1711.00595}
  {arXiv:1711.00595} \BibitemShut {NoStop}%
\bibitem [{\citenamefont {Burkard}\ \emph {et~al.}(1999)\citenamefont
  {Burkard}, \citenamefont {Loss},\ and\ \citenamefont
  {DiVincenzo}}]{Burkard1999}%
  \BibitemOpen
  \bibfield  {author} {\bibinfo {author} {\bibfnamefont {G.}~\bibnamefont
  {Burkard}}, \bibinfo {author} {\bibfnamefont {D.}~\bibnamefont {Loss}}, \
  and\ \bibinfo {author} {\bibfnamefont {D.~P.}\ \bibnamefont {DiVincenzo}},\
  }\href {\doibase 10.1103/PhysRevB.59.2070} {\bibfield  {journal} {\bibinfo
  {journal} {Physical Review B}\ }\textbf {\bibinfo {volume} {59}},\ \bibinfo
  {pages} {2070} (\bibinfo {year} {1999})}\BibitemShut {NoStop}%
\bibitem [{\citenamefont {Li}\ \emph {et~al.}(2010)\citenamefont {Li},
  \citenamefont {Cywi{\'{n}}ski}, \citenamefont {Culcer}, \citenamefont {Hu},\
  and\ \citenamefont {{Das Sarma}}}]{Li2009}%
  \BibitemOpen
  \bibfield  {author} {\bibinfo {author} {\bibfnamefont {Q.}~\bibnamefont
  {Li}}, \bibinfo {author} {\bibfnamefont {{\L}.}~\bibnamefont
  {Cywi{\'{n}}ski}}, \bibinfo {author} {\bibfnamefont {D.}~\bibnamefont
  {Culcer}}, \bibinfo {author} {\bibfnamefont {X.}~\bibnamefont {Hu}}, \ and\
  \bibinfo {author} {\bibfnamefont {S.}~\bibnamefont {{Das Sarma}}},\ }\href
  {\doibase 10.1103/PhysRevB.81.085313} {\bibfield  {journal} {\bibinfo
  {journal} {Physical Review B}\ }\textbf {\bibinfo {volume} {81}},\ \bibinfo
  {pages} {085313} (\bibinfo {year} {2010})}\BibitemShut {NoStop}%
\bibitem [{\citenamefont {Mehl}\ \emph {et~al.}(2014)\citenamefont {Mehl},
  \citenamefont {Bluhm},\ and\ \citenamefont {DiVincenzo}}]{Mehl2014c}%
  \BibitemOpen
  \bibfield  {author} {\bibinfo {author} {\bibfnamefont {S.}~\bibnamefont
  {Mehl}}, \bibinfo {author} {\bibfnamefont {H.}~\bibnamefont {Bluhm}}, \ and\
  \bibinfo {author} {\bibfnamefont {D.~P.}\ \bibnamefont {DiVincenzo}},\ }\href
  {\doibase 10.1103/PhysRevB.90.045404} {\bibfield  {journal} {\bibinfo
  {journal} {Physical Review B}\ }\textbf {\bibinfo {volume} {90}},\ \bibinfo
  {pages} {045404} (\bibinfo {year} {2014})}\BibitemShut {NoStop}%
\bibitem [{\citenamefont {Baart}\ \emph {et~al.}(2016)\citenamefont {Baart},
  \citenamefont {Fujita}, \citenamefont {Reichl}, \citenamefont {Wegscheider},\
  and\ \citenamefont {Vandersypen}}]{Baart2016a}%
  \BibitemOpen
  \bibfield  {author} {\bibinfo {author} {\bibfnamefont {T.~A.}\ \bibnamefont
  {Baart}}, \bibinfo {author} {\bibfnamefont {T.}~\bibnamefont {Fujita}},
  \bibinfo {author} {\bibfnamefont {C.}~\bibnamefont {Reichl}}, \bibinfo
  {author} {\bibfnamefont {W.}~\bibnamefont {Wegscheider}}, \ and\ \bibinfo
  {author} {\bibfnamefont {L.~M.~K.}\ \bibnamefont {Vandersypen}},\ }\href
  {\doibase 10.1038/nnano.2016.188} {\bibfield  {journal} {\bibinfo  {journal}
  {Nature Nanotechnology}\ }\textbf {\bibinfo {volume} {12}},\ \bibinfo {pages}
  {26} (\bibinfo {year} {2016})}\BibitemShut {NoStop}%
\bibitem [{\citenamefont {Mi}\ \emph {et~al.}(2017)\citenamefont {Mi},
  \citenamefont {Benito}, \citenamefont {Putz}, \citenamefont {Zajac},
  \citenamefont {Taylor}, \citenamefont {Burkard},\ and\ \citenamefont
  {Petta}}]{Mi2017}%
  \BibitemOpen
  \bibfield  {author} {\bibinfo {author} {\bibfnamefont {X.}~\bibnamefont
  {Mi}}, \bibinfo {author} {\bibfnamefont {M.}~\bibnamefont {Benito}}, \bibinfo
  {author} {\bibfnamefont {S.}~\bibnamefont {Putz}}, \bibinfo {author}
  {\bibfnamefont {D.~M.}\ \bibnamefont {Zajac}}, \bibinfo {author}
  {\bibfnamefont {J.~M.}\ \bibnamefont {Taylor}}, \bibinfo {author}
  {\bibfnamefont {G.}~\bibnamefont {Burkard}}, \ and\ \bibinfo {author}
  {\bibfnamefont {J.~R.}\ \bibnamefont {Petta}},\ }\href
  {http://arxiv.org/abs/1710.03265} {\ ,\ \bibinfo {pages} {1} (\bibinfo {year}
  {2017})},\ \Eprint {http://arxiv.org/abs/1710.03265} {arXiv:1710.03265}
  \BibitemShut {NoStop}%
\bibitem [{\citenamefont {Srinivasa}\ \emph {et~al.}(2015)\citenamefont
  {Srinivasa}, \citenamefont {Xu},\ and\ \citenamefont
  {Taylor}}]{Srinivasa2015}%
  \BibitemOpen
  \bibfield  {author} {\bibinfo {author} {\bibfnamefont {V.}~\bibnamefont
  {Srinivasa}}, \bibinfo {author} {\bibfnamefont {H.}~\bibnamefont {Xu}}, \
  and\ \bibinfo {author} {\bibfnamefont {J.~M.}\ \bibnamefont {Taylor}},\
  }\href {\doibase 10.1103/PhysRevLett.114.226803} {\bibfield  {journal}
  {\bibinfo  {journal} {Physical Review Letters}\ }\textbf {\bibinfo {volume}
  {114}},\ \bibinfo {pages} {226803} (\bibinfo {year} {2015})}\BibitemShut
  {NoStop}%
\bibitem [{\citenamefont {Croot}\ \emph {et~al.}(2017)\citenamefont {Croot},
  \citenamefont {Pauka}, \citenamefont {Watson}, \citenamefont {Gardner},
  \citenamefont {Fallahi}, \citenamefont {Manfra},\ and\ \citenamefont
  {Reilly}}]{Croot2017}%
  \BibitemOpen
  \bibfield  {author} {\bibinfo {author} {\bibfnamefont {X.~G.}\ \bibnamefont
  {Croot}}, \bibinfo {author} {\bibfnamefont {S.~J.}\ \bibnamefont {Pauka}},
  \bibinfo {author} {\bibfnamefont {J.~D.}\ \bibnamefont {Watson}}, \bibinfo
  {author} {\bibfnamefont {G.~C.}\ \bibnamefont {Gardner}}, \bibinfo {author}
  {\bibfnamefont {S.}~\bibnamefont {Fallahi}}, \bibinfo {author} {\bibfnamefont
  {M.~J.}\ \bibnamefont {Manfra}}, \ and\ \bibinfo {author} {\bibfnamefont
  {D.~J.}\ \bibnamefont {Reilly}},\ }\href {http://arxiv.org/abs/1707.06479}
  {\bibfield  {journal} {\bibinfo  {journal} {arXiv preprint}\ ,\ \bibinfo
  {pages} {1}} (\bibinfo {year} {2017})},\ \Eprint
  {http://arxiv.org/abs/1707.06479} {arXiv:1707.06479} \BibitemShut {NoStop}%
\bibitem [{\citenamefont {Malinowski}(2017)}]{Malinowski_Thesis}%
  \BibitemOpen
  \bibfield  {author} {\bibinfo {author} {\bibfnamefont {F.~K.}\ \bibnamefont
  {Malinowski}},\ }\href@noop {} {\emph {\bibinfo {title} {Noise suppression
  and long-range exchange coupling for gallium arsenide spin qubits}}}\
  (\bibinfo  {publisher} {Ph.D. Thesis, Center for Quantum Devices, Niels Bohr
  Institute, University of Copenhagen},\ \bibinfo {year} {2017})\
  Chap.~\bibinfo {chapter} {14}\BibitemShut {NoStop}%
\bibitem [{\citenamefont {Martins}\ \emph {et~al.}(2017)\citenamefont
  {Martins}, \citenamefont {Malinowski}, \citenamefont {Nissen}, \citenamefont
  {Fallahi}, \citenamefont {Gardner}, \citenamefont {Manfra}, \citenamefont
  {Marcus},\ and\ \citenamefont {Kuemmeth}}]{Martins2017}%
  \BibitemOpen
  \bibfield  {author} {\bibinfo {author} {\bibfnamefont {F.}~\bibnamefont
  {Martins}}, \bibinfo {author} {\bibfnamefont {F.~K.}\ \bibnamefont
  {Malinowski}}, \bibinfo {author} {\bibfnamefont {P.~D.}\ \bibnamefont
  {Nissen}}, \bibinfo {author} {\bibfnamefont {S.}~\bibnamefont {Fallahi}},
  \bibinfo {author} {\bibfnamefont {G.~C.}\ \bibnamefont {Gardner}}, \bibinfo
  {author} {\bibfnamefont {M.~J.}\ \bibnamefont {Manfra}}, \bibinfo {author}
  {\bibfnamefont {C.~M.}\ \bibnamefont {Marcus}}, \ and\ \bibinfo {author}
  {\bibfnamefont {F.}~\bibnamefont {Kuemmeth}},\ }\href {\doibase
  10.1103/PhysRevLett.119.227701} {\bibfield  {journal} {\bibinfo  {journal}
  {Physical Review Letters}\ }\textbf {\bibinfo {volume} {119}},\ \bibinfo
  {pages} {227701} (\bibinfo {year} {2017})}\BibitemShut {NoStop}%
\bibitem [{\citenamefont {Malinowski}\ \emph {et~al.}(2018)\citenamefont
  {Malinowski}, \citenamefont {Martins}, \citenamefont {Smith}, \citenamefont
  {Bartlett}, \citenamefont {Doherty}, \citenamefont {Nissen}, \citenamefont
  {Fallahi}, \citenamefont {Gardner}, \citenamefont {Manfra}, \citenamefont
  {Marcus},\ and\ \citenamefont {Kuemmeth}}]{Malinowski2017a}%
  \BibitemOpen
  \bibfield  {author} {\bibinfo {author} {\bibfnamefont {F.~K.}\ \bibnamefont
  {Malinowski}}, \bibinfo {author} {\bibfnamefont {F.}~\bibnamefont {Martins}},
  \bibinfo {author} {\bibfnamefont {T.~B.}\ \bibnamefont {Smith}}, \bibinfo
  {author} {\bibfnamefont {S.~D.}\ \bibnamefont {Bartlett}}, \bibinfo {author}
  {\bibfnamefont {A.~C.}\ \bibnamefont {Doherty}}, \bibinfo {author}
  {\bibfnamefont {P.~D.}\ \bibnamefont {Nissen}}, \bibinfo {author}
  {\bibfnamefont {S.}~\bibnamefont {Fallahi}}, \bibinfo {author} {\bibfnamefont
  {G.~C.}\ \bibnamefont {Gardner}}, \bibinfo {author} {\bibfnamefont {M.~J.}\
  \bibnamefont {Manfra}}, \bibinfo {author} {\bibfnamefont {C.~M.}\
  \bibnamefont {Marcus}}, \ and\ \bibinfo {author} {\bibfnamefont
  {F.}~\bibnamefont {Kuemmeth}},\ }\href {\doibase 10.1103/PhysRevX.8.011045}
  {\bibfield  {journal} {\bibinfo  {journal} {Phys. Rev. X}\ }\textbf {\bibinfo
  {volume} {8}},\ \bibinfo {pages} {011045} (\bibinfo {year}
  {2018})}\BibitemShut {NoStop}%
\bibitem [{\citenamefont {Wang}\ \emph {et~al.}(2012)\citenamefont {Wang},
  \citenamefont {Bishop}, \citenamefont {Kestner}, \citenamefont {Barnes},
  \citenamefont {Sun},\ and\ \citenamefont {{Das Sarma}}}]{Wang2012}%
  \BibitemOpen
  \bibfield  {author} {\bibinfo {author} {\bibfnamefont {X.}~\bibnamefont
  {Wang}}, \bibinfo {author} {\bibfnamefont {L.~S.}\ \bibnamefont {Bishop}},
  \bibinfo {author} {\bibfnamefont {J.}~\bibnamefont {Kestner}}, \bibinfo
  {author} {\bibfnamefont {E.}~\bibnamefont {Barnes}}, \bibinfo {author}
  {\bibfnamefont {K.}~\bibnamefont {Sun}}, \ and\ \bibinfo {author}
  {\bibfnamefont {S.}~\bibnamefont {{Das Sarma}}},\ }\href {\doibase
  10.1038/ncomms2003} {\bibfield  {journal} {\bibinfo  {journal} {Nature
  Communications}\ }\textbf {\bibinfo {volume} {3}},\ \bibinfo {pages} {997}
  (\bibinfo {year} {2012})}\BibitemShut {NoStop}%
\bibitem [{\citenamefont {Kestner}\ \emph {et~al.}(2013)\citenamefont
  {Kestner}, \citenamefont {Wang}, \citenamefont {Bishop}, \citenamefont
  {Barnes},\ and\ \citenamefont {{Das Sarma}}}]{Kestner2013}%
  \BibitemOpen
  \bibfield  {author} {\bibinfo {author} {\bibfnamefont {J.~P.}\ \bibnamefont
  {Kestner}}, \bibinfo {author} {\bibfnamefont {X.}~\bibnamefont {Wang}},
  \bibinfo {author} {\bibfnamefont {L.~S.}\ \bibnamefont {Bishop}}, \bibinfo
  {author} {\bibfnamefont {E.}~\bibnamefont {Barnes}}, \ and\ \bibinfo {author}
  {\bibfnamefont {S.}~\bibnamefont {{Das Sarma}}},\ }\href {\doibase
  10.1103/PhysRevLett.110.140502} {\bibfield  {journal} {\bibinfo  {journal}
  {Physical Review Letters}\ }\textbf {\bibinfo {volume} {110}},\ \bibinfo
  {pages} {140502} (\bibinfo {year} {2013})}\BibitemShut {NoStop}%
\bibitem [{\citenamefont {Wang}\ \emph {et~al.}(2014)\citenamefont {Wang},
  \citenamefont {Bishop}, \citenamefont {Barnes}, \citenamefont {Kestner},\
  and\ \citenamefont {{Das Sarma}}}]{Wang2014}%
  \BibitemOpen
  \bibfield  {author} {\bibinfo {author} {\bibfnamefont {X.}~\bibnamefont
  {Wang}}, \bibinfo {author} {\bibfnamefont {L.~S.}\ \bibnamefont {Bishop}},
  \bibinfo {author} {\bibfnamefont {E.}~\bibnamefont {Barnes}}, \bibinfo
  {author} {\bibfnamefont {J.~P.}\ \bibnamefont {Kestner}}, \ and\ \bibinfo
  {author} {\bibfnamefont {S.}~\bibnamefont {{Das Sarma}}},\ }\href {\doibase
  10.1103/PhysRevA.89.022310} {\bibfield  {journal} {\bibinfo  {journal}
  {Physical Review A}\ }\textbf {\bibinfo {volume} {89}},\ \bibinfo {pages}
  {022310} (\bibinfo {year} {2014})}\BibitemShut {NoStop}%
\bibitem [{\citenamefont {Goelman}\ \emph {et~al.}(1989)\citenamefont
  {Goelman}, \citenamefont {Vega},\ and\ \citenamefont {Zax}}]{Goelman1989}%
  \BibitemOpen
  \bibfield  {author} {\bibinfo {author} {\bibfnamefont {G.}~\bibnamefont
  {Goelman}}, \bibinfo {author} {\bibfnamefont {S.}~\bibnamefont {Vega}}, \
  and\ \bibinfo {author} {\bibfnamefont {D.}~\bibnamefont {Zax}},\ }\href
  {\doibase 10.1016/0022-2364(89)90077-2} {\bibfield  {journal} {\bibinfo
  {journal} {Journal of Magnetic Resonance (1969)}\ }\textbf {\bibinfo {volume}
  {81}},\ \bibinfo {pages} {423} (\bibinfo {year} {1989})}\BibitemShut
  {NoStop}%
\bibitem [{\citenamefont {Wagner}\ \emph {et~al.}(1992)\citenamefont {Wagner},
  \citenamefont {Merkt},\ and\ \citenamefont {Chaplik}}]{Wagner1992}%
  \BibitemOpen
  \bibfield  {author} {\bibinfo {author} {\bibfnamefont {M.}~\bibnamefont
  {Wagner}}, \bibinfo {author} {\bibfnamefont {U.}~\bibnamefont {Merkt}}, \
  and\ \bibinfo {author} {\bibfnamefont {A.~V.}\ \bibnamefont {Chaplik}},\
  }\href {\doibase 10.1103/PhysRevB.45.1951} {\bibfield  {journal} {\bibinfo
  {journal} {Physical Review B}\ }\textbf {\bibinfo {volume} {45}},\ \bibinfo
  {pages} {1951} (\bibinfo {year} {1992})}\BibitemShut {NoStop}%
\bibitem [{\citenamefont {Baruffa}\ \emph {et~al.}(2010)\citenamefont
  {Baruffa}, \citenamefont {Stano},\ and\ \citenamefont
  {Fabian}}]{Baruffa2010a}%
  \BibitemOpen
  \bibfield  {author} {\bibinfo {author} {\bibfnamefont {F.}~\bibnamefont
  {Baruffa}}, \bibinfo {author} {\bibfnamefont {P.}~\bibnamefont {Stano}}, \
  and\ \bibinfo {author} {\bibfnamefont {J.}~\bibnamefont {Fabian}},\ }\href
  {\doibase 10.1103/PhysRevB.82.045311} {\bibfield  {journal} {\bibinfo
  {journal} {Physical Review B}\ }\textbf {\bibinfo {volume} {82}},\ \bibinfo
  {pages} {045311} (\bibinfo {year} {2010})}\BibitemShut {NoStop}%
\bibitem [{\citenamefont {Zumb{\"{u}}hl}\ \emph {et~al.}(2004)\citenamefont
  {Zumb{\"{u}}hl}, \citenamefont {Marcus}, \citenamefont {Hanson},\ and\
  \citenamefont {Gossard}}]{Zumbuhl2004}%
  \BibitemOpen
  \bibfield  {author} {\bibinfo {author} {\bibfnamefont {D.~M.}\ \bibnamefont
  {Zumb{\"{u}}hl}}, \bibinfo {author} {\bibfnamefont {C.~M.}\ \bibnamefont
  {Marcus}}, \bibinfo {author} {\bibfnamefont {M.~P.}\ \bibnamefont {Hanson}},
  \ and\ \bibinfo {author} {\bibfnamefont {A.~C.}\ \bibnamefont {Gossard}},\
  }\href {\doibase 10.1103/PhysRevLett.93.256801} {\bibfield  {journal}
  {\bibinfo  {journal} {Physical Review Letters}\ }\textbf {\bibinfo {volume}
  {93}},\ \bibinfo {pages} {256801} (\bibinfo {year} {2004})}\BibitemShut
  {NoStop}%
\bibitem [{\citenamefont {Mehl}\ and\ \citenamefont
  {DiVincenzo}(2014)}]{Mehl2014a}%
  \BibitemOpen
  \bibfield  {author} {\bibinfo {author} {\bibfnamefont {S.}~\bibnamefont
  {Mehl}}\ and\ \bibinfo {author} {\bibfnamefont {D.~P.}\ \bibnamefont
  {DiVincenzo}},\ }\href {\doibase 10.1103/PhysRevB.90.195424} {\bibfield
  {journal} {\bibinfo  {journal} {Physical Review B}\ }\textbf {\bibinfo
  {volume} {90}},\ \bibinfo {pages} {195424} (\bibinfo {year}
  {2014})}\BibitemShut {NoStop}%
\bibitem [{\citenamefont {Lieb}\ and\ \citenamefont {Mattis}(1962)}]{Lieb1962}%
  \BibitemOpen
  \bibfield  {author} {\bibinfo {author} {\bibfnamefont {E.}~\bibnamefont
  {Lieb}}\ and\ \bibinfo {author} {\bibfnamefont {D.}~\bibnamefont {Mattis}},\
  }\href {\doibase 10.1103/PhysRev.125.164} {\bibfield  {journal} {\bibinfo
  {journal} {Physical Review}\ }\textbf {\bibinfo {volume} {125}},\ \bibinfo
  {pages} {164} (\bibinfo {year} {1962})}\BibitemShut {NoStop}%
\bibitem [{\citenamefont {Feynman}(1998)}]{2electron_theorem}%
  \BibitemOpen
  \bibfield  {author} {\bibinfo {author} {\bibfnamefont {R.~P.}\ \bibnamefont
  {Feynman}},\ }\href@noop {} {\emph {\bibinfo {title} {Statistical mechanics:
  A set of lectures}}}\ (\bibinfo  {publisher} {Westview Press},\ \bibinfo
  {year} {1998})\ Chap.\ \bibinfo {chapter} {11.3}\BibitemShut {NoStop}%
\bibitem [{\citenamefont {Riiser}\ and\ \citenamefont
  {Ravndal}(1993)}]{Riiser1993a}%
  \BibitemOpen
  \bibfield  {author} {\bibinfo {author} {\bibfnamefont {A.~C.}\ \bibnamefont
  {Riiser}}\ and\ \bibinfo {author} {\bibfnamefont {F.}~\bibnamefont
  {Ravndal}},\ }\href {\doibase 10.1103/PhysRevB.48.5648} {\bibfield  {journal}
  {\bibinfo  {journal} {Phys. Rev. B}\ }\textbf {\bibinfo {volume} {48}},\
  \bibinfo {pages} {5648} (\bibinfo {year} {1993})}\BibitemShut {NoStop}%
\bibitem [{\citenamefont {Folk}\ \emph {et~al.}(2001)\citenamefont {Folk},
  \citenamefont {Marcus}, \citenamefont {Berkovits}, \citenamefont {Kurland},
  \citenamefont {Aleiner},\ and\ \citenamefont {Altshuler}}]{Folk2001}%
  \BibitemOpen
  \bibfield  {author} {\bibinfo {author} {\bibfnamefont {J.~A.}\ \bibnamefont
  {Folk}}, \bibinfo {author} {\bibfnamefont {C.~M.}\ \bibnamefont {Marcus}},
  \bibinfo {author} {\bibfnamefont {R.}~\bibnamefont {Berkovits}}, \bibinfo
  {author} {\bibfnamefont {I.~L.}\ \bibnamefont {Kurland}}, \bibinfo {author}
  {\bibfnamefont {I.~L.}\ \bibnamefont {Aleiner}}, \ and\ \bibinfo {author}
  {\bibfnamefont {B.~L.}\ \bibnamefont {Altshuler}},\ }\href {\doibase
  10.1238/Physica.Topical.090a00026} {\bibfield  {journal} {\bibinfo  {journal}
  {Physica Scripta}\ }\textbf {\bibinfo {volume} {T90}},\ \bibinfo {pages} {26}
  (\bibinfo {year} {2001})}\BibitemShut {NoStop}%
\bibitem [{\citenamefont {Lindemann}\ \emph {et~al.}(2002)\citenamefont
  {Lindemann}, \citenamefont {Ihn}, \citenamefont {Heinzel}, \citenamefont
  {Zwerger}, \citenamefont {Ensslin}, \citenamefont {Maranowski},\ and\
  \citenamefont {Gossard}}]{Lindemann2002}%
  \BibitemOpen
  \bibfield  {author} {\bibinfo {author} {\bibfnamefont {S.}~\bibnamefont
  {Lindemann}}, \bibinfo {author} {\bibfnamefont {T.}~\bibnamefont {Ihn}},
  \bibinfo {author} {\bibfnamefont {T.}~\bibnamefont {Heinzel}}, \bibinfo
  {author} {\bibfnamefont {W.}~\bibnamefont {Zwerger}}, \bibinfo {author}
  {\bibfnamefont {K.}~\bibnamefont {Ensslin}}, \bibinfo {author} {\bibfnamefont
  {K.}~\bibnamefont {Maranowski}}, \ and\ \bibinfo {author} {\bibfnamefont
  {A.~C.}\ \bibnamefont {Gossard}},\ }\href {\doibase
  10.1103/PhysRevB.66.195314} {\bibfield  {journal} {\bibinfo  {journal}
  {Physical Review B}\ }\textbf {\bibinfo {volume} {66}},\ \bibinfo {pages}
  {195314} (\bibinfo {year} {2002})}\BibitemShut {NoStop}%
\bibitem [{\citenamefont {Boyd}(1984)}]{Boyd1984}%
  \BibitemOpen
  \bibfield  {author} {\bibinfo {author} {\bibfnamefont {R.~J.}\ \bibnamefont
  {Boyd}},\ }\href {\doibase 10.1038/310480a0} {\bibfield  {journal} {\bibinfo
  {journal} {Nature}\ }\textbf {\bibinfo {volume} {310}},\ \bibinfo {pages}
  {480} (\bibinfo {year} {1984})}\BibitemShut {NoStop}%
\bibitem [{\citenamefont {Altland}\ and\ \citenamefont
  {Simons}(2010)}]{CMFT_Alexander}%
  \BibitemOpen
  \bibfield  {author} {\bibinfo {author} {\bibfnamefont {A.}~\bibnamefont
  {Altland}}\ and\ \bibinfo {author} {\bibfnamefont {B.~D.}\ \bibnamefont
  {Simons}},\ }\href@noop {} {\emph {\bibinfo {title} {Condensed matter field
  theory}}}\ (\bibinfo  {publisher} {Cambridge University Press},\ \bibinfo
  {year} {2010})\ pp.\ \bibinfo {pages} {59--61}\BibitemShut {NoStop}%
\bibitem [{\citenamefont {Durand}\ and\ \citenamefont
  {Barthelat}(1975)}]{Durand1975}%
  \BibitemOpen
  \bibfield  {author} {\bibinfo {author} {\bibfnamefont {P.}~\bibnamefont
  {Durand}}\ and\ \bibinfo {author} {\bibfnamefont {J.-C.}\ \bibnamefont
  {Barthelat}},\ }\href {\doibase 10.1007/BF00963468} {\bibfield  {journal}
  {\bibinfo  {journal} {Theoretica Chimica Acta}\ }\textbf {\bibinfo {volume}
  {38}},\ \bibinfo {pages} {283} (\bibinfo {year} {1975})}\BibitemShut
  {NoStop}%
\bibitem [{\citenamefont {Malrieu}\ \emph {et~al.}(2014)\citenamefont
  {Malrieu}, \citenamefont {Caballol}, \citenamefont {Calzado}, \citenamefont
  {de~Graaf},\ and\ \citenamefont {Guih{\'{e}}ry}}]{Malrieu2014}%
  \BibitemOpen
  \bibfield  {author} {\bibinfo {author} {\bibfnamefont {J.~P.}\ \bibnamefont
  {Malrieu}}, \bibinfo {author} {\bibfnamefont {R.}~\bibnamefont {Caballol}},
  \bibinfo {author} {\bibfnamefont {C.~J.}\ \bibnamefont {Calzado}}, \bibinfo
  {author} {\bibfnamefont {C.}~\bibnamefont {de~Graaf}}, \ and\ \bibinfo
  {author} {\bibfnamefont {N.}~\bibnamefont {Guih{\'{e}}ry}},\ }\href {\doibase
  10.1021/cr300500z} {\bibfield  {journal} {\bibinfo  {journal} {Chemical
  Reviews}\ }\textbf {\bibinfo {volume} {114}},\ \bibinfo {pages} {429}
  (\bibinfo {year} {2014})}\BibitemShut {NoStop}%
\bibitem [{\citenamefont {Barnes}\ \emph {et~al.}(2011)\citenamefont {Barnes},
  \citenamefont {Kestner}, \citenamefont {Nguyen},\ and\ \citenamefont {{Das
  Sarma}}}]{Barnes2011}%
  \BibitemOpen
  \bibfield  {author} {\bibinfo {author} {\bibfnamefont {E.}~\bibnamefont
  {Barnes}}, \bibinfo {author} {\bibfnamefont {J.~P.}\ \bibnamefont {Kestner}},
  \bibinfo {author} {\bibfnamefont {N.~T.~T.}\ \bibnamefont {Nguyen}}, \ and\
  \bibinfo {author} {\bibfnamefont {S.}~\bibnamefont {{Das Sarma}}},\ }\href
  {\doibase 10.1103/PhysRevB.84.235309} {\bibfield  {journal} {\bibinfo
  {journal} {Physical Review B}\ }\textbf {\bibinfo {volume} {84}},\ \bibinfo
  {pages} {235309} (\bibinfo {year} {2011})}\BibitemShut {NoStop}%
\bibitem [{\citenamefont {Yang}\ \emph {et~al.}(2013)\citenamefont {Yang},
  \citenamefont {Rossi}, \citenamefont {Ruskov}, \citenamefont {Lai},
  \citenamefont {Mohiyaddin}, \citenamefont {Lee}, \citenamefont {Tahan},
  \citenamefont {Klimeck}, \citenamefont {Morello},\ and\ \citenamefont
  {Dzurak}}]{Yang2013}%
  \BibitemOpen
  \bibfield  {author} {\bibinfo {author} {\bibfnamefont {C.~H.}\ \bibnamefont
  {Yang}}, \bibinfo {author} {\bibfnamefont {A.}~\bibnamefont {Rossi}},
  \bibinfo {author} {\bibfnamefont {R.}~\bibnamefont {Ruskov}}, \bibinfo
  {author} {\bibfnamefont {N.~S.}\ \bibnamefont {Lai}}, \bibinfo {author}
  {\bibfnamefont {F.~A.}\ \bibnamefont {Mohiyaddin}}, \bibinfo {author}
  {\bibfnamefont {S.}~\bibnamefont {Lee}}, \bibinfo {author} {\bibfnamefont
  {C.}~\bibnamefont {Tahan}}, \bibinfo {author} {\bibfnamefont
  {G.}~\bibnamefont {Klimeck}}, \bibinfo {author} {\bibfnamefont
  {A.}~\bibnamefont {Morello}}, \ and\ \bibinfo {author} {\bibfnamefont
  {A.~S.}\ \bibnamefont {Dzurak}},\ }\href {\doibase 10.1038/ncomms3069}
  {\bibfield  {journal} {\bibinfo  {journal} {Nat. Commun.}\ }\textbf {\bibinfo
  {volume} {4}},\ \bibinfo {pages} {1} (\bibinfo {year} {2013})}\BibitemShut
  {NoStop}%
\bibitem [{\citenamefont {Reimann}\ and\ \citenamefont
  {Manninen}(2002)}]{Reimann2002}%
  \BibitemOpen
  \bibfield  {author} {\bibinfo {author} {\bibfnamefont {S.~M.}\ \bibnamefont
  {Reimann}}\ and\ \bibinfo {author} {\bibfnamefont {M.}~\bibnamefont
  {Manninen}},\ }\href {\doibase 10.1103/RevModPhys.74.1283} {\bibfield
  {journal} {\bibinfo  {journal} {Rev. Mod. Phys.}\ }\textbf {\bibinfo {volume}
  {74}},\ \bibinfo {pages} {1283} (\bibinfo {year} {2002})}\BibitemShut
  {NoStop}%
\bibitem [{\citenamefont {Nielsen}\ \emph {et~al.}(2013)\citenamefont
  {Nielsen}, \citenamefont {Barnes}, \citenamefont {Kestner},\ and\
  \citenamefont {{Das Sarma}}}]{Nielsen2013a}%
  \BibitemOpen
  \bibfield  {author} {\bibinfo {author} {\bibfnamefont {E.}~\bibnamefont
  {Nielsen}}, \bibinfo {author} {\bibfnamefont {E.}~\bibnamefont {Barnes}},
  \bibinfo {author} {\bibfnamefont {J.~P.}\ \bibnamefont {Kestner}}, \ and\
  \bibinfo {author} {\bibfnamefont {S.}~\bibnamefont {{Das Sarma}}},\ }\href
  {\doibase 10.1103/PhysRevB.88.195131} {\bibfield  {journal} {\bibinfo
  {journal} {Physical Review B}\ }\textbf {\bibinfo {volume} {88}},\ \bibinfo
  {pages} {195131} (\bibinfo {year} {2013})}\BibitemShut {NoStop}%
\bibitem [{\citenamefont {Higginbotham}\ \emph {et~al.}(2014)\citenamefont
  {Higginbotham}, \citenamefont {Kuemmeth}, \citenamefont {Hanson},
  \citenamefont {Gossard},\ and\ \citenamefont {Marcus}}]{Higginbotham2014}%
  \BibitemOpen
  \bibfield  {author} {\bibinfo {author} {\bibfnamefont {A.~P.}\ \bibnamefont
  {Higginbotham}}, \bibinfo {author} {\bibfnamefont {F.}~\bibnamefont
  {Kuemmeth}}, \bibinfo {author} {\bibfnamefont {M.~P.}\ \bibnamefont
  {Hanson}}, \bibinfo {author} {\bibfnamefont {A.~C.}\ \bibnamefont {Gossard}},
  \ and\ \bibinfo {author} {\bibfnamefont {C.~M.}\ \bibnamefont {Marcus}},\
  }\href {\doibase 10.1103/PhysRevLett.112.026801} {\bibfield  {journal}
  {\bibinfo  {journal} {Physical Review Letters}\ }\textbf {\bibinfo {volume}
  {112}},\ \bibinfo {pages} {026801} (\bibinfo {year} {2014})}\BibitemShut
  {NoStop}%
\bibitem [{\citenamefont {Hada}\ and\ \citenamefont {Eto}(2003)}]{Hada2003}%
  \BibitemOpen
  \bibfield  {author} {\bibinfo {author} {\bibfnamefont {Y.}~\bibnamefont
  {Hada}}\ and\ \bibinfo {author} {\bibfnamefont {M.}~\bibnamefont {Eto}},\
  }\href {\doibase 10.1103/PhysRevB.68.155322} {\bibfield  {journal} {\bibinfo
  {journal} {Phys. Rev. B}\ }\textbf {\bibinfo {volume} {68}},\ \bibinfo
  {pages} {155322} (\bibinfo {year} {2003})}\BibitemShut {NoStop}%
\bibitem [{\citenamefont {{A linear geometry was the most practical choice to
  study the minimum conditions under which negative exchange energy occurs in a
  multielectron quantum dot. Alternatively}}()}]{Note}%
  \BibitemOpen
  \bibfield  {author} {\bibinfo {author} {\bibnamefont {{A linear geometry was
  the most practical choice to study the minimum conditions under which
  negative exchange energy occurs in a multielectron quantum dot.
  Alternatively}}},\ }\href@noop {} {\bibinfo  {journal} {{ a triangular system
  would have been more general from a theoretical point of view and could have
  led to interesting physics. However, the inclusion of an extra tunnel
  coupling would bring in additional effects such as superexchange that would
  likely obscure the negative exchange phenomenon that we are targeting in our
  analysis. Moreover, a system of three quantum dots with tunnel coupling
  between every pair of dots is not desirable from the point of view of quantum
  computing architectures, where square lattices are needed to implement
  quantum error correcting surface codes}}\ }\BibitemShut {NoStop}%
\end{thebibliography}%

\end{document}